\documentclass[pra,onecolumn,showpacs,nofootinbib,preprint]{revtex4-1}

\usepackage{amsmath,amsfonts,amssymb,bm}
\usepackage{dcolumn}
\usepackage[final]{graphicx}
\usepackage{bm}
\usepackage{comment}
\usepackage{dsfont}
\usepackage{float}
\usepackage{rotating}
\usepackage{leftidx}
\usepackage{bbold}
\usepackage{color}

\newcommand{\be}{\begin{equation}}
\newcommand{\ee}{\end{equation}}
\newcommand{\bea}{\begin{eqnarray}}
\newcommand{\eea}{\end{eqnarray}}
\newcommand{\bit}{\begin{itemize}}
\newcommand{\eit}{\end{itemize}}

\newcommand{\bfl}{\begin{flushright}}
\newcommand{\efl}{\end{flushright}}

\newcommand{\non}{\nonumber \\}
\newcommand{\nonu}{\nonumber}

\newcommand{\ra}{\rangle}
\newcommand{\la}{\langle}

\newcommand{\mus}{\mathbf{u}}
\newcommand{\mvs}{\mathbf{v}}

\newcommand{\mr}{\mathbf{r}}

\renewcommand\Re{\operatorname{Re}}
\renewcommand\Im{\operatorname{Im}}

\includecomment{pdffigure}

\begin{document}

\bibliographystyle{apsrev4-1}

\title{Excitation spectra of fragmented condensates by linear response:
General theory and application to a condensate in a double-well potential}

\author{Julian Grond}
\thanks{Corresponding author: julian.grond@pci.uni-heidelberg.de}
\author{Alexej I. Streltsov}
\author{Lorenz S. Cederbaum}
\affiliation{Theoretische Chemie, Physikalisch--Chemisches Institut, Universit\"at Heidelberg,   Germany}

\author{Ofir E. Alon}
\affiliation{Department of Physics, University of Haifa at Oranim, Tivon 36006, Israel}

\date{\today}

\begin{abstract}

Linear response of simple (i.e., condensed) Bose-Einstein condensates is known to lead to the Bogoliubov-
de Gennes equations. Here, we derive linear response for fragmented Bose-Einstein
condensates, i.e., for the case where the many-body wave function is not a product of one, but
of several single-particle states (orbitals). This gives one access to excitation spectra and
response amplitudes of systems beyond the Gross-Pitaevskii description. Our approach is
based on the number-conserving variational time-dependent mean field theory [O. E. Alon,
A. I. Streltsov, and L. S. Cederbaum, Phys. Lett. A \textbf{362}, 453 (2007)], which describes the time
evolution of best-mean field states. Correspondingly, we call our linear response theory for
fragmented states LR-BMF. In the derivation it follows naturally that excitations are
orthogonal to the ground-state orbitals. As applications excitation spectra of Bose-Einstein
condensates in double-well potentials are calculated. Both symmetric and asymmetric double-wells are studied for several interaction strengths and barrier heights. The cases of condensed
and two-fold fragmented ground states are compared. Interestingly, even in such situations
where the response frequencies of the two cases are computed to be close to each other, which
is the situation for the excitations well below the barrier, striking differences in the density
response in momentum space are found. For excitations with an energy of the order of the barrier height, both the energies
and the density response of condensed and fragmented systems are very different. In
fragmented systems there is a class of ``swapped" excitations where an atom is transfered to
the neighboring well. The mechanism of its origin is discussed. In asymmetric wells, the response of a fragmented system is purely
local (i.e., finite in either one or the other well) with different frequencies for the left and right fragments. This finding is in stark
contrast to that for condensed systems.

\end{abstract}



\maketitle 
 
\section{Introduction}

The precision with which ultra-cold bosonic quantum gases and Bose-Einstein condensates (BECs) can be manipulated nowadays \cite{folman:02,grimm:00} has allowed to study not only their ground states but also excited states and dynamics, such as dynamical splitting of a BEC \cite{schumm:05}, Josephson dynamics \cite{albiez:05,levy:07}, dynamical creation of number squeezed states \cite{esteve:08,maussang:10}, quantum optimal control \cite{buecker:11}, and multi-band physics \cite{will:10,mark:11}.
  
Crucial for the understanding of the dynamics of quantum gases are excitation spectra. Theoretically, they have been widely studied for BECs using the standard Bogoliubov-de Gennes (BdG) equations, which can be regarded as the linear response of the ground state described by the Gross-Pitaevskii (GP) equation \cite{bogoliubov:47,dalfovo:99,leggett:01,pethick:08} to an external perturbation. Predicted spectra \cite{edwards:96,perez:96,stringari:96b} compared accurately to experiments measuring for example collective excitations of trapped BECs \cite{jin:96,mewes:96}, the speed of sound propagation of quasi-particles \cite{andrews:97}, and excitations of BECs in the bulk regime \cite{steinhauer:01,ozeri:05}. Similar theoretical and experimental studies have been performed for two-species BECs, modeled by two coupled Gross-Pitaevskii equations (2GP) and the corresponding Bogoliubov-de Gennes equations \cite{myatt:97,Esry.prl:97,Esry.pra:98,Pu:98,sorensen:02}.

In those works it is assumed that the atoms are essentially condensed. Therefore, the ground state of the system is well described by the GP (or 2GP) equation. However, there are numerous examples where this is not the case and, instead, one deals with fragmented condensates \cite{nozieres:82,nozieres:96,streltsov:06,mueller:06,sakmann:08}. Examples involve condensates in symmetric \cite{spekkens:99,javanainen:99,menotti:01,streltsov:07} and asymmetric double-well potentials \cite{cederbaum:03,alon.prl2:05,streltsov:06}. Mott insulator states in few-well systems \cite{orzel:01,streltsov2:04} and optical lattices \cite{jaksch:98,greiner:02} represent multiple fragmented states. Other examples of fragmentation can be found in cold atom systems exhibiting translational and rotational symmetry \cite{mueller:06,tsatsos:10}, in attractive condensates \cite{wilkin:97,leggett:99,alon.epl:04,cederbaum.prl:08,streltsov.prl:11}, in low dimensions \cite{kolomeisky:00,petrov:00}, for long-range interactions \cite{bader:09}, and in metastable situations \cite{shin.prl2:04, streltsov1:04}. In optical lattice systems, unusual depletion \cite{xu:06} and excitation frequencies measured  in the presence of an external harmonic potential \cite{tuchman:06} could not be explained within Bogoliubov theory. In all those situations, where the ground state is fragmented, the standard BdG approach for calculating excitation spectra is not applicable. 
 
A theory capable of describing statics of fragmentation phenomena is the \emph{best-mean field} (BMF) method \cite{cederbaum:03}. This method is based on a variational framework and a general mean-field ansatz for the state. It has been successfully applied to evaluate the ground-state fragmentation of BECs in double-trap potentials \cite{streltsov2:04} and allowed to identify stable fragmented excited states in repulsive condensates with energies below the GP self-consistent excited states \cite{streltsov1:04}. The pathway from condensation to fermionization, passing a variety of fragmented states, has been demonstrated in \cite{alon.prl2:05}. Similarly a variety of new Mott-insulator phases has been found for optical lattices \cite{alon.prl1:05}. This method has been extended to bosonic mixtures, showing interesting demixing scenarios \cite{alon.prl:06}.  A dynamical theory based on a similar ansatz as the BMF is provided by the \emph{time-dependent multi-orbital mean field} (TDMF) \cite{alon.pla:07} method. For example, it allowed to predict interaction-induced \emph{self-interference} fringes \cite{cederbaum.prl:07}. 

In this paper, we will generalize the standard BdG equations originally derived for condensed (or simple \cite{leggett:01}) BECs to the case of fragmented BECs. In particular, linear response of the TDMF will provide a tool for studying excitation spectra and properties of the excitations of fragmented condensates. The derived equations are general and can thus be employed for the calculation of spectra of systems with an arbitrary degree of fragmentation.

As an application of our response theory for fragmented BECs we will investigate Bose-Einstein condensates in symmetric and asymmetric double-well potentials. Those are prototypical systems which are best suited to test the theory, because they show fragmentation even for moderate interactions. Typically, double-well potentials have been studied either with a classical two-mode description \cite{giovanazzi:00, raghavan:99} (`Josephson physics'), or within a two-mode Bose-Hubbard model \cite{milburn:97,javanainen:99}. Other studies suggested that more than one state per well has to be taken into account \cite{menotti:01,dounas:07}. Recent dynamical studies \cite{sakmann:09,grond.pra:09,grond.pra:09b,sakmann.pra:10,grond.njp:11} involve the \emph{multiconfigurational time-dependent Hartree for Bosons} (MCTDHB) method \cite{streltsov:07,alon:08}.  A few excited states have been calculated in \cite{masiello:05,streltsov:06} using self-consistent methods. Bogoliubov approximation has been applied to few-site or lattice models \cite{paraoanu:01,oosten:01}, and 
to double-well BECs 
\cite{japha:11}. Linear response studies beyond BdG have been performed so far with the sine-Gordon model \cite{gritsev:07} or within Gutzwiller-approximation \cite{menotti:08,krut:10,snoek:11}. However, a thorough study of excitations is still missing and will be provided in this work. In particular we compare excitation energies and the density response of condensed and two-fold fragmented states in double-well potentials.

The structure of the paper is as follows. First, in Sec.~\ref{sec:the} we introduce the underlying theoretical tools and models such as the standard BdG equations in Sec.~\ref{subsec:lin_gp}, and the TDMF method in Sec.~\ref{subsec:tim}.  In Sec.~\ref{sec:lin_frag} we present the derivation of the linear response theory for fragmented BECs, discuss its properties and provide expressions for important observables.  Thereafter, in Sec.~\ref{sec:app} we apply the derived equations to calculate excitation spectra of BECs in double-well potentials. First, we discuss the possible structures of the ground states in this system, and then present our linear response results for symmetric and asymmetric double-well potentials in  Secs.~\ref{subsec:sym} and \ref{subsec:asy}, respectively. Finally we summarize in Sec.~\ref{sec:con} our findings and draw conclusions. There are also three appendices. Appendix~\ref{app:lin} deals with some algebraic subtleties of the linear response equations. 
Special cases of linear response relevant for the application part are discussed in Appendix~\ref{app:spec}, where we also give explicitly the corresponding response matrices. In Appendix~\ref{app:com} we compare qualitatively the linear response of a two-fold fragmented system to that of a two-species system of distinct BECs. Finally, in Appendix~\ref{app:lr_bh} we derive the linear response of the two-site Bose-Hubbard model, which allows to quantify the importance of hopping excitations.
 
\section{Theoretical concepts \label{sec:the}}

A system of $N$ interacting atoms in an external potential is described by the many-body Hamiltonian \cite{dalfovo:99,leggett:01}:
\begin{equation}\label{eq:ham}
 \hat H= \sum_{i=1}^N\hat h(\mathbf{r}_i)+\lambda_0\sum_{i>j=1}^N \delta(\mathbf{r}_i-\mathbf{r}_j)\,,
\end{equation}
with the single-particle Hamiltonian\footnote{We work in dimensionless units where the energy is measured in terms of $\frac{\hbar^2}{mL^2}$. We choose $\hbar=1$, set the mass of a $^{87}$Rb atom to one, and the unit of length to $L=1\mu m$. This gives units of energy and time $E=h\cdot 116.26$ Hz and $T=1.37$ ms, respectively.}
\be
\hat h(\mathbf{r})=-\boldsymbol{\nabla}^2/2+V(\mathbf{r})\,.
\ee
 $N$ is the total number of atoms. The first term on the right-hand side of Eq.~\eqref{eq:ham} describes kinetic and potential energy. The second term accounts for atom-atom interactions with interaction parameter $\lambda_0$, which is proportional to the s-wave scattering length \cite{dalfovo:99,leggett:01}. We use the commonly employed delta potential, but stress that the following formulas and derivations do not rely on the type of interaction potential.

\subsection{Linear response of the Gross-Pitaevskii equation \label{subsec:lin_gp}}

The standard method for calculating excitation spectra of interacting bosons at zero temperature is solving the Bogoliubov-de Gennes equations \cite{dalfovo:99,leggett:01,pethick:08}. They can be derived as the linear response of the GP equation to an external time oscillating potential \cite{edwards:96,ruprecht:96}.  The linearized equations of motion are equivalent to the equations obtained  when treating the many-body Hamiltonian, Eq.~\eqref{eq:ham}, in Bogoliubov approximation \cite{bogoliubov:47,gardiner:97,castin:98}, or in the random-phase approximation (RPA) \cite{esry:97}.  All those approaches assume that the system is essentially condensed, i.e., one eigenvalue dominates the one-body reduced density matrix of the system \cite{penrose:56}. The non-condensate fraction has to be much smaller than unity in the Bogoliubov or RPA treatments.


We shortly sketch here the derivation of the BdG equations \cite{ruprecht:96}. The GP equation, which assumes that all atoms reside in a single orbital, reads
\be\label{eq:GP1}
i\dot{\phi}=\hat H_{GP}\phi,\quad \hat H_{GP}=\hat h+\lambda|\phi|^2\,,
\ee
with the interaction strength $\lambda=\lambda_0(N-1)$.
 A small time-dependent periodic perturbation of the external potential, $\hat h(\mathbf{r})\rightarrow\hat h(\mathbf{r})+\delta\hat{h}(\mathbf{r},t)$, can be written generally as:  
\be\label{eq:pert}
\delta\hat{h}(\mathbf{r},t)=f^+(\mathbf{r})e^{-i\omega t}+f^-(\mathbf{r})e^{i\omega t}\,,
\ee
with the probe frequency $\omega$ and the amplitudes $f^{\pm}$ real.\footnote{We note that without a perturbation, the following procedure amounts to linearizing the equations of motion Eq.~\eqref{eq:GP1}. However, in order to show that the spectrum defined by the linearized equations indeed corresponds to the frequencies of the excitation energies, we derive explicitly the response to a small perturbation. The exact shapes of the perturbations $f^+(\mathbf{r})$ and $f^-(\mathbf{r})$ do not influence the linear response spectrum. Rather the pole strength of the various excitations in the perturbed orbitals is affected.} Making the ansatz
\be\label{eq:GP_orb}
\sqrt{N}\phi(\mathbf{r},t)=e^{-i\mu t}\left[\sqrt{N}\phi^0(\mathbf{r})+u(\mathbf{r}) e^{-i\omega t} +v^*(\mathbf{r}) e^{i\omega t}\right]\,,
\ee 
as an expansion around the solution of the static GP equation $\phi^0(\mathbf{r})$ (with chemical potential $\mu$) and 
with small amplitudes $|u\ra$ and $|v\ra$, one arrives at the equation
\be\label{eq:lr_gp}
\left(\boldsymbol{\mathcal{L}_{BdG}}-\omega\right)\left(\begin{array}{c}|u\ra \\ |v\ra\end{array}\right)=\left(\begin{array}{c}-\sqrt{N}f^+|\phi^0\ra \\ \sqrt{N}f^-|\phi^{0,*}\ra\end{array}\right)\,.
\ee
The linear response matrix reads
\be\label{eq:lrm_gp}
\boldsymbol{\mathcal{L}_{BdG}}=\left(\begin{array}{cc} \hat H_{GP} +\lambda|\phi^0|^2-\mu & \lambda(\phi^0)^2\\
-\lambda(\phi^{0,*})^{2} & -(\hat H_{GP} +\lambda|\phi^0|^2-\mu)
 \end{array}\right)\,.
\ee
Eq.~\eqref{eq:lr_gp} with the right hand side equal to zero is referred to as  Bogoliubov-de Gennes equations. They determine the \emph{response frequencies} $\omega_k$, and also the \emph{response amplitudes} $(|u^k\ra,|v^k\ra)^T$.
Using them we can solve the linear response Eq.~\eqref{eq:lr_gp} for $(|u\ra,|v\ra)^T$ with respect to a given perturbation. Inserted into the ansatz, Eq.~\eqref{eq:GP_orb}, we obtain finally
\be\label{eq:response_gp}
\phi(\mathbf{r},t)=e^{-i\mu t}\left\{\phi^0(\mathbf{r})+\frac{1}{\sqrt{N}}\sum_k\left[ \gamma_k u^k(\mathbf{r}) e^{-i\omega t}+\gamma_k^* v^{k,*}(\mathbf{r}) e^{i\omega t}\right]/(\omega-\omega_k)\right\}\,,
\ee
 with the response weights (or pole strengths)\footnote{The response diverges at the resonance frequencies, which seems to be unphysical. When performing a time-dependent simulation, however, the response is damped due to effects beyond the linear regime, which reestablishes the physically expected behavior \cite{ruprecht:96}.}
 \be
 \gamma_k=\sqrt{N}\int d\mathbf{r}[u^{k,*}(\mathbf{r})f^+(\mathbf{r})\phi^0(\mathbf{r})+v^{k,*}(\mathbf{r})f^-(\mathbf{r})\phi^{0,*}(\mathbf{r})]\,.
 \ee
   In deriving the linear response theory for fragmented BECs we will keep the above nomenclature as much as possible.
 
\subsection{Time-dependent multi-orbital mean field \label{subsec:tim}}

The Hilbert space of the many-body Schr\"odinger equation with Hamiltonian given in Eq.~\eqref{eq:ham} is huge for the atom numbers one is interested in and which are typically used in experiments (say $N\gtrsim 100$). The following variational framework provides an efficient method to numerically solve this equation \cite{cederbaum:03,alon.pla:07}. The starting point is an ansatz for an arbitrarily fragmented mean-field state in terms of time-dependent orbitals: 
\be\label{eq:MFansatz}
\Psi(\mathbf{r}_1,...,\mathbf{r}_N,t)=\hat{\mathcal{S}}\phi_1(\mathbf{r}_1,t)...\phi_1(\mathbf{r}_{n_1},t)\phi_2(\mathbf{r}_{n_1+1},t)...\phi_2(\mathbf{r}_{n_1+n_2},t)...\phi_M(\mathbf{r}_N,t)\,.
\ee
Here, we put $n_1$ atoms into orbital $1$, $n_2$ atoms into orbital $2$, ..., and $n_M$ into orbital $M$, with $\sum_{l=1}^M n_l=N$. $\hat{\mathcal{S}}$ is the symmetrization operator for bosons. Obviously, a GP state, where all atoms occupy one and the same orbital, is a special case of Eq.~\eqref{eq:MFansatz}, i.e., for $M=1$. We are thus about to generalize the GP equation.  

The ansatz Eq.~\eqref{eq:MFansatz} is now used to formulate an action functional
\be\label{eq:action}
S=\int dt\left\{\la\Psi|\hat H-i\frac{\partial}{\partial t}|\Psi\ra-\sum_{i,j}^M n_i\mu_{ij}(t)[\la\phi_i|\phi_j\ra-\delta_{ij}]\right\}\,,
\ee
where $\mu_{ij}(t)$ are Lagrange multipliers which ensure the orthonormality of the time-dependent orbitals $\phi_i(\mathbf{r},t)$. The first expression under the integral in Eq.~\eqref{eq:action} is evaluated to be
\be
\la\Psi|\hat H-i\frac{\partial}{\partial t}|\Psi\ra=\sum_{i=1}^M n_i\Biggl[h_{ii}-\left(i\frac{\partial}{\partial t}\right)_{ii}+\lambda_0\frac{n_i-1}{2}W_{iiii}+\sum_{j\neq i}^M\lambda_0 n_j W_{ijij}\Biggr]\,,
\ee
with the matrix elements
\be
h_{ii}=\int\phi_i^*(\mathbf{r},t)\hat h(\mathbf{r},t)\phi_i(\mathbf{r},t)d\mathbf{r}\,,\quad
\left(i\frac{\partial}{\partial t}\right)_{ii}=i\int \phi_i^*(\mathbf{r},t)\dot{\phi}_i(\mathbf{r},t)d\mathbf{r}\,,
\ee
and
\be
W_{ijij}=\int|\phi_i(\mathbf{r},t)|^2|\phi_j(\mathbf{r},t)|^2 d\mathbf{r}\,.
\ee
The variational principle now requires the stationarity of the action, Eq.~\eqref{eq:action}, with respect to the orbitals:
\be
\frac{\delta S}{\delta\phi_i^*(\mathbf{r},t)}=0\,,\quad i=1,...,M\,.
\ee
From this we obtain after some algebra the TDMF equations for a chosen occupation $(n_1,n_2,...,n_M)$
\be\label{eq:TDMFP}
\hat{P}i|\dot{\phi}_i\ra=\hat{P}\left[\hat h+\lambda_0(n_i-1)|\phi_i|^2+\sum_{j\neq i}^{M}2\lambda_0n_j|\phi_j|^2\right]|\phi_i\ra\,,\quad i=1,...,M\,.
\ee
The projector $\hat{P}=\mathbb{1}-\sum_{s=1}^{M}|\phi_s\ra\la\phi_s|$, resulting from the Lagrange multipliers,  keeps the orbitals orthonormal throughout the time propagation. The energy of the time-dependent mean-field states as obtained from Eq.~\eqref{eq:TDMFP} is conserved  whenever the Lagrange multipliers\footnote{Note that we use a different definition of the Lagrange multipliers as compared to Ref.~\cite{alon.pla:07}. The convention which is used there can be obtained by replacing $\mu_{ij}(t)\rightarrow \mu_{ij}(t)/n_i$.} are hermitian, i.e., $n_i\mu_{ij}(t)=n_j\mu_{ji}^*(t)$, or alternatively if $\la\phi_i|\dot{\phi}_j\ra=0$, ($\,i,j=1,...,M$). Since there is no rigorous proof for the hermicity of $\mu_{ij}(t)$ throughout the propagation in time, one enforces the latter condition as an additional constraint. As a consequence, the projector on the left-hand side of Eq.~\eqref{eq:TDMFP} can be omitted. This has also the positive effect to simplify those integro-differential equations. With this we arrive at the final form of the TDMF equations:
\be\label{eq:TDMF}
i|\dot{\phi}_i\ra=\hat{P}\left[\hat h+\lambda_0(n_i-1)|\phi_i|^2+\sum_{j\neq i}^{M}2\lambda_0n_j|\phi_j|^2\right]|\phi_i\ra\,,\quad i=1,...,M\,.
\ee
In this work we find that the linear response of both the full form Eq.~\eqref{eq:TDMFP} and the working equations [see Eq.~\eqref{eq:TDMF}] give rise to the same excitation energies and response amplitudes, and, therefore, to the same perturbed orbitals $\phi_i(\mathbf{r},t)$. Algebraic subtleties of the linear response of the full form are discussed in Appendix~\ref{app:lin}.

\section{Linear response theory for fragmented Bose-Einstein condensates \label{sec:lin_frag}}

\subsection{Derivation \label{sec:der}}

We now derive the linear response theory for fragmented BECs. For this purpose we add a time-dependent perturbation to the TDMF equations [see Eq.~\eqref{eq:TDMF}], in a way as it has been done in Eq.~\eqref{eq:pert} for the GP equation. The corresponding equations of motion can be written in a compact form as ($i=1,...,M$):
\be\label{eq:basiseq}
\left(\hat Z_i-i\frac{\partial}{\partial t}\right)|\phi_i\ra-\sum_{j=1}^{M}\mu_{ij}(t)|\phi_j\ra=-\delta \hat h(t)|\phi_i\ra\,,
\ee
where
\bea\label{eq:ham_TDMF}
\hat Z_i&=&\hat h+\lambda_0(n_i-1)|\phi_i|^2 + \sum_{j\neq i}^{M}2\lambda_0 n_j|\phi_j|^2
\eea
and
\be
\mu_{ij}(t)=\la\phi_j|\hat Z_{i}+\delta\hat h(t)|\phi_{i}\ra\,.
\ee
The Lagrange multipliers $\mu_{ij}(t)$ account for the orthonormality of the orbitals $\phi_i$, analogously as the projector $\hat P$ in the TDMF equations [see Eq.~\eqref{eq:TDMF}]. Keeping $\mu_{ij}(t)$ explicitly helps to identify throughout the derivation the terms which originate from the orthogonalization with respect to the orbitals.   We expand the perturbed orbitals around stationary solutions as 
 \be\label{eq:ansatz0}
 \phi_i(\mathbf{r},t)\approx\phi_i^0(\mathbf{r})+\delta\phi_i(\mathbf{r},t)\,.
 \ee
By keeping terms up to first order in $\delta\phi_i(\mathbf{r},t)$ and $f^{\pm}(\mathbf{r})$ we obtain:
\bea\label{eq:LR1}
&&\left(\hat Z_i^0-i\frac{\partial}{\partial t}+\lambda_0(n_i-1)|\phi_i^0|^2\right)|\delta\phi_i\ra+\sum_{j\neq i}^M 2\lambda_0 n_j\phi_j^{0,*}\phi_i^0|\delta\phi_j\ra+\lambda_0(n_i-1)(\phi_i^0)^2|\delta\phi_i^{*}\ra\non
&&+\sum_{j\neq i}^M 2\lambda_0 n_j\phi_j^0\phi_i^0|\delta\phi_j^{*}\ra-\sum_{j=1}^{M}\left[\mu_{ij}^0|\delta\phi_j\ra+\delta \mu_{ij}(t)|\phi_j^0\ra\right]=-\delta \hat h(t)|\phi_i^0\ra\,.
\eea
The zeroth-order contribution $\hat Z_i^0$ is defined as in Eq.~\eqref{eq:ham_TDMF} but with $\phi_i\rightarrow\phi_i^0$. Without the perturbation, i.e., $\delta\hat h=0$, Eq.~\eqref{eq:basiseq} is solved by the time-independent orbitals $\phi_i^0(\mathbf{r})$\footnote{We note that due to the presence of the Lagrange multipliers, the stationary solutions of TDMF carry no time-dependent phase factors as they do in Eq.~\eqref{eq:GP_orb}.}, which are solutions of the \emph{best-mean field} equations \cite{cederbaum:03}. Those equations describe the stationary states of the TDMF\footnote{Strictly speaking, the best-mean field is defined as the optimal orbitals at the energetically optimal occupation.}:
\be\label{eq:bmf}
Z_{i}^0|\phi_i^0\ra = \sum_{j=1}^M \mu_{ij}^0|\phi_j^0\ra\,.\\
\ee
In many cases linear response is performed for ground-state orbitals, but $\phi_i^0(\mr)$ could be excited stationary orbitals as well.  The Lagrange multipliers to zeroth order are given as $\mu_{ij}^0=\la\phi_{j}^0|\hat Z_{i}^0|\phi_{i}^0\ra$.  The perturbed Lagrange multipliers are evaluated to be
\bea\label{eq:muij}
\delta\mu_{ij}(t)&=&\delta\left[\la\phi_j|\left(\hat Z_{i}+\delta\hat h(t)\right)|\phi_{i}\ra\right]\\
&=&\sum_{l=1}^M\mu_{il}^0\la\delta\phi_j|\phi_{l}^0\ra+\la\phi_j^0|\delta\Bigl(\hat Z_{i}|\phi_{i}\ra\Bigr)+\la\phi_j^0|\delta\hat h(t)|\phi_{i}^0\ra\,.\nonu
\eea
In order to arrive at the first term we used that the unperturbed orbitals $\phi_{j}^0(\mr)$ fulfill the best-mean field equations [see Eq.~\eqref{eq:bmf}]. 
Essentially, the matrix elements $\delta\mu_{ij}(t)$ lead to the same projectors $\hat P$ as in TDMF, acting on Eq.~\eqref{eq:LR1}. This is directly obvious for all terms of $\delta\mu_{ij}(t)$ except for the first one. Using integration by parts and exchanging the indices of the summations, we can  rewrite the first term of the sum $\sum_{j=1}^M\delta\mu_{ij}(t)|\phi_j^0(\mr)\ra$   as
\be
\sum_{j,l=1}^M\mu_{il}^0\la\delta\phi_j|\phi_{l}^0\ra|\phi_{j}^0\ra=-\sum_{j=1}^M\mu_{ij}^0(1-\hat P)|\delta\phi_{j}\ra\,.
\ee
We see that it corresponds to a projector on the term of Eq.~\eqref{eq:LR1} which is proportional to the Lagrange multipliers $\mu_{ij}^0$. With this we find
\bea\label{eq:LR2}
&&\hat P\Biggl[\left(\hat Z_i^0+\lambda_0(n_i-1)|\phi_i^0|^2\right)|\delta\phi_i\ra-\sum_{j=1}^{M}\mu_{ij}^0|\delta\phi_j\ra+\sum_{j\neq i}^M 2\lambda_0 n_j\phi_j^{0,*}\phi_i^0|\delta\phi_j\ra\non
&&+\lambda_0(n_i-1)(\phi_i^0)^2|\delta\phi_i^{*}\ra+\sum_{j\neq i}^M 2\lambda_0 n_j\phi_j^0\phi_i^0|\delta\phi_j^{*}\ra\Biggr]-i\frac{\partial}{\partial t}|\delta\phi_i\ra=-\hat P\delta \hat h(t)|\phi_i^0\ra\,.
\eea
This equation has, similarly to the TDMF equations [see Eq.~\eqref{eq:TDMF}], projectors on all terms except for the time derivative. 

By using the ansatz
\be\label{eq:ansatz_uv}
\sqrt{n_i}\delta\phi_i(\mathbf{r},t)=u_i(\mathbf{r}) e^{-i\omega t} +v_i^*(\mathbf{r}) e^{i\omega t}
\ee
 for the time-dependent perturbation to the orbitals ($\omega$ is the probe frequency), and by equating like powers of $e^{\pm i\omega t}$, we obtain the linear response system 
\bea\label{eq:lr}
&&\left(\boldsymbol{\mathcal{P}}\boldsymbol{\mathcal{L}}-\omega\right)\left(\begin{array}{c}|\mathbf{u}\ra \\ |\mathbf{v}\ra\end{array}\right)=\boldsymbol{\mathcal{P}}\left(\begin{array}{c}-f^+|\boldsymbol{\phi^0_n}\ra \\ f^{-}|\boldsymbol{\phi^{0,*}_n}\ra\end{array}\right)\,,
\eea
where $f^{\pm}$ are the real amplitudes of the external perturbation. We switched to a vector notation in order to have a compact representation of the multi-orbital problem. For example, we denote the vector of stationary orbitals, multiplied by the square root of the population $n_i$ for each orbital, as $|\boldsymbol{\phi^0_n}\ra=(|\sqrt{n_1}\phi_1^0\ra,|\sqrt{n_2}\phi_2^0\ra,...,|\sqrt{n_M}\phi_M^0\ra)^T$. 
$\boldsymbol{\mathcal{L}}$ is the linear response matrix, with
\be\label{eq:lrm}
\boldsymbol{\mathcal{L}}=\left(\begin{array}{cc} \boldsymbol{Z^0}-\boldsymbol{\mu^0}+\boldsymbol{A} & \boldsymbol{B}\\
-\boldsymbol{B}^* & -(\boldsymbol{Z^{0}}-\boldsymbol{\mu^{0,*}})-\boldsymbol{A}^*
 \end{array}\right)\,.\ee
Here, $\boldsymbol{Z^0}$ is a diagonal matrix containing the $\hat{Z}_i^0$.  We group the matrix elements of the diagonal contributions originating from atom-atom interactions in $\boldsymbol{A}$, as well as the off-diagonal ones in $\boldsymbol{B}$. They are given as
\bea
&&A_{ij}=\Biggl\{\begin{array}{l} \lambda_0(n_i-1)|\phi_i^0|^2,\quad i=j\\2\lambda_0 \sqrt{n_i n_j}\phi_j^{0,*}\phi_i^0,\quad i\neq j \end{array}\,,\non
&&B_{ij}=\Biggl\{\begin{array}{l} \lambda_0(n_i-1)(\phi_i^0)^2,\quad i=j\\ 2\lambda_0 \sqrt{n_i n_j}\phi_j^0\phi_i^0,\quad i\neq j \end{array}\,.
\eea 
 The projector matrix contains twice as many projectors as the number of orbitals $M$ ($i,j=1,...,2M$)
\be
\mathcal{P}_{ij}=\left\{\begin{array}{l} \hat P\,,\quad\mathrm{for}\quad i=j\le M \\ \hat P^*\,,\quad\mathrm{for}\quad i=j> M \\ 0\,,\quad(i\neq j)\end{array}\right.\;,
\ee
where $\hat{P}^*=\mathbb{1}-\sum_{s=1}^{M}|\phi_s^*\ra\la\phi_s^*|$. By acting with $\left(1-\boldsymbol{\mathcal{P}}\right)$ on the linear response system~\eqref{eq:lr}, we find that for $|\omega|>0$ the solution is orthogonal to the stationary orbitals $\phi_i^0(\mr)$, i.e.,
\be\label{eq:condP}
\boldsymbol{\mathcal{P}}\left(\begin{array}{c}|\mathbf{u}\ra \\ |\mathbf{v}\ra\end{array}\right)=\left(\begin{array}{c}|\mathbf{u}\ra \\ |\mathbf{v}\ra\end{array}\right)\,.
\ee
This allows us to add an additional projector in Eq.~\eqref{eq:lr} (i.e., replacing $\boldsymbol{\mathcal{P}}\boldsymbol{\mathcal{L}}\rightarrow\boldsymbol{\mathcal{P}}\boldsymbol{\mathcal{L}}\boldsymbol{\mathcal{P}}$). In order to find the excitation energies $\omega^k$ in Eq.~\eqref{eq:lr} one has to solve the eigenvalue problem
\be\label{eq:evL}
\boldsymbol{\mathcal{P}}\boldsymbol{\mathcal{L}}\boldsymbol{\mathcal{P}}\left(\begin{array}{c}|\mathbf{u}^k\ra\\|\mathbf{v}^k\ra\end{array}\right)=\omega^k\left(\begin{array}{c}|\mathbf{u}^k\ra\\|\mathbf{v}^k\ra\end{array}\right)\,.
\ee
Most importantly and as a consequence of the projectors, for $|\omega^k|>0$ each component of the eigenvectors  $(\mathbf{u}^{k},\mathbf{v}^{k})^T$ is orthogonal to the stationary orbitals $\phi_i^0$. 

We call the linear response Eq.~\eqref{eq:evL}, together with the linear response matrix given in Eq.~\eqref{eq:lrm},  LR-BMF. The special case $M=1$ is referred to as LR-GP, which is the linear response of the number-conserving version of the GP equation (for the differences to the linear response matrix of BdG see Appendix~\ref{app:spec}).

In order to find the orthonormalization relations of the response amplitudes for non-zero eigenvalues of Eq.~\eqref{eq:evL}, we study the symmetries of $\boldsymbol{\mathcal{P}}\boldsymbol{\mathcal{L}}\boldsymbol{\mathcal{P}}$ similar as in Ref.~\cite{castin:98}. 
First, a time-reversal spin-flip-like symmetry:
\be\label{eq:sigma1}
\boldsymbol{\Sigma_1}\boldsymbol{\mathcal{P}}\boldsymbol{\mathcal{L}}\boldsymbol{\mathcal{P}}\boldsymbol{\Sigma_1}=-\boldsymbol{\mathcal{P}}^*\boldsymbol{\mathcal{L}}^*\boldsymbol{\mathcal{P}}^*\,,
\ee
where the matrix $[\Sigma_1]_{ij}=\delta_{i,j-M}+\delta_{i-M,j}$ ($i,j=1,...,2M$) permutes the $i$-th and the $M+i$-th raws, just as the first Pauli matrix $\sigma_1=\bigl(\begin{smallmatrix}0&1\\1&0\end{smallmatrix}\bigr)$ does for $M=1$. Further, we have
\be\label{eq:sigma3}
\boldsymbol{\Sigma_3}\boldsymbol{\mathcal{P}}\boldsymbol{\mathcal{L}}\boldsymbol{\mathcal{P}}\boldsymbol{\Sigma_3}=\left( \boldsymbol{\mathcal{P}}\boldsymbol{\mathcal{L}}\boldsymbol{\mathcal{P}} \right)^{\dagger}\,,
\ee
where the matrix 
\be
[\Sigma_3]_{ij}=\Biggl\{\begin{array}{ll}\delta_{i,j}&,\quad\mathrm{for}\quad i,j\le M\\ -\delta_{i,j}&,\quad\mathrm{for}\quad i,j> M \end{array}\;.
\ee
For the case $M=1$ this reduces to the third Pauli matrix $\sigma_3=\bigl(\begin{smallmatrix}1&0\\0&-1\end{smallmatrix}\bigr)$. From Eq.~\eqref{eq:sigma1} we learn that, whenever $(|\mathbf u^k\ra,|\mathbf v^k\ra)^T$ is an eigenvector of $\boldsymbol{\mathcal{P}}\boldsymbol{\mathcal{L}}\boldsymbol{\mathcal{P}}$ with eigenvalue $\omega^k$, then $(|\mathbf v^{k,*}\ra,|\mathbf u^{k,*}\ra)^T$ is an eigenvector with eigenvalue $-(\omega^{k})^*$. From Eq.~\eqref{eq:sigma3} we find that $(|\mathbf u^k\ra,-|\mathbf v^k\ra)^T$ is an eigenvector of $\left( \boldsymbol{\mathcal{P}}\boldsymbol{\mathcal{L}}\boldsymbol{\mathcal{P}} \right)^{\dagger}$ with eigenvalue $\omega^k$, which allows us to construct the adjoint basis. From those considerations follow the biorthonormalization relations 
\bea\label{eq:orth}
&&\la \mathbf u^k|\mathbf u^{k'}\ra-\la \mathbf v^k|\mathbf v^{k'}\ra=\delta_{kk'}\,,\non
&&\la \mathbf v^k|\mathbf u^{k',*}\ra-\la \mathbf u^k|\mathbf v^{k',*}\ra=0\,.
\eea

It is obvious that all eigenvectors of $\boldsymbol{\mathcal{P}}\boldsymbol{\mathcal{L}}\boldsymbol{\mathcal{P}}$ for $\omega_k=0$ lie in the space spanned by the stationary orbitals of the unperturbed problem. The completeness relation then reads
\bea
\mathbb{1}&=&\sum_{i,j=1}^M\left(\begin{array}{c}|\mathbf{u}_{ij}^0\ra \\ 0 \end{array}\right)\left(\la\mathbf{u}_{ij}^0|, 0 \right)+\sum_{i,j=1}^M\left(\begin{array}{c}0\\|\mathbf{v}_{ij}^0\ra  \end{array}\right)\left(0,\la\mathbf{v}_{ij}^0| \right)+\sum_{k>0}\left(\begin{array}{c}|\mathbf{u}^k\ra \\ |\mathbf{v}^k\ra\end{array}\right)\left(\la\mathbf{u}^k| , -\la\mathbf{v}^k|\right)\non
&&+\sum_{k>0}\left(\begin{array}{c}|\mathbf{v}^{k,*}\ra \\ |\mathbf{u}^{k,*}\ra\end{array}\right)\left(\la\mathbf{v}^{k,*}| , -\la\mathbf{u}^{k,*}|\right)\,.
\eea
$\left(|\mus^k\ra,|\mvs^k\ra\right)^T$ are the eigenvectors with strictly positive eigenvalues $\omega^k>0$.\footnote{We found numerically real response frequencies $\omega^k$ when starting from stationary orbitals.} The $i$-th element of $|\mathbf{u}_{ij}^0\ra$ ($|\mathbf{v}_{ij}^0\ra$) is  equal to $\phi_j^0$ ($\phi_j^{0,*}$) ($i,j=1,...,M$). All other elements of $|\mathbf{u}_{ij}^0\ra$ and $|\mathbf{v}_{ij}^0\ra$ vanish. 

Now we solve Eq.~\eqref{eq:lr} by expanding the response vectors  as well as the perturbation with the eigenvectors of $\boldsymbol{\mathcal{P}}\boldsymbol{\mathcal{L}}\boldsymbol{\mathcal{P}}$ orthogonal to the stationary orbitals $\phi_i^0(\mr)$. 
The ansatz for the response amplitudes then reads
\be\label{eq:ansatz1}
\left(\begin{array}{c}|\mathbf u \ra\\ |\mathbf v\ra\end{array}\right)=\sum_{k} c_k\left(\begin{array}{c}|\mus^k \ra\\ |\mvs^k\ra\end{array}\right)\,,
\ee
and for the perturbation
\be\label{eq:ansatz2}
-\boldsymbol{\mathcal{P}}\left(\begin{array}{l}-f^+(\mathbf{r})|\boldsymbol{\phi^0_n}\ra \\ f^{-}(\mathbf{r})|\boldsymbol{\phi^{0,*}_n}\ra\end{array}\right)=\sum_{k} \gamma_k\left(\begin{array}{c}|\mus^k\ra \\ |\mvs^k\ra\end{array}\right)\,.
\ee
Now $c_k$ and $\gamma_k$ have to be determined. Substituting Eqs.~\eqref{eq:ansatz1} and \eqref{eq:ansatz2} into Eq.~\eqref{eq:lr}, we obtain
\bea\label{eq:coeff_comp_det}
&&\sum_k c_k(\omega_k-\omega)\left(\begin{array}{c}|\mus^k\ra \\ |\mvs^k\ra\end{array}\right)=-\sum_{k} \gamma_k\left(\begin{array}{c}|\mus^k\ra \\ |\mvs^k\ra\end{array}\right)\,,
\eea
where $\omega_k$ is defined in Eq.~\eqref{eq:evL}. From comparing coefficients in Eq.~\eqref{eq:coeff_comp_det} we get an expression for the coefficients $c_k$. 
Inserted in Eq.~\eqref{eq:ansatz1} this leads to a solution for the response amplitudes of the form
\be
\left(\begin{array}{c}|\mathbf u\ra \\ |\mathbf v\ra\end{array}\right)=\sum_k\frac{\gamma_k}{\omega-\omega_k}\left(\begin{array}{c}|\mus^k\ra \\ |\mvs^k\ra\end{array}\right)\,.
\ee
Reinserting the amplitudes into the ansatz for the orbitals, Eqs.~\eqref{eq:ansatz0} and \eqref{eq:ansatz_uv}, we arrive at the final solution for the time-dependent orbitals in linear response ($i=1,...,M$):
\be\label{eq:response}
\phi_i(\mathbf{r},t)=\phi_i^0(\mathbf{r})+\sum_k\frac{1}{\sqrt{n_i}}\left[ \gamma_k u_i^k e^{-i\omega t}+\gamma_k^* v_i^{k,*} e^{i\omega t}\right]/(\omega-\omega_k)\,.
\ee
Thus the orbitals, and with them the density, show the largest response at frequencies $\omega_k$. Moreover, the response for a fixed frequency $\omega_k$ is not necessarily equally strong for all the orbitals. This is because  the components of the response amplitudes $u_j^k$ and $v_j^k$  are not normalized,  but rather the whole amplitude vector [see Eq.~\eqref{eq:orth}]. The response weights, which quantify the intensity of the response, are given as
\be\label{eq:weight}
\gamma_k=\sum_{j=1}^M\sqrt{n_j}\int d\mathbf{r}\left[u_j^{k,*}(\mathbf{r})f^+(\mathbf{r})\phi_j^0(\mathbf{r})+v_j^{k,*}(\mathbf{r})f^-(\mathbf{r})\phi_j^{0,*}(\mathbf{r})\right]\,.
\ee
Similarly as the orbitals [Eq.~\eqref{eq:response}], it is dominated by the largest components of the response amplitudes.

\subsection{Density oscillations}

When probing the linear response through a time-dependent perturbation, an observable quantity is the oscillation of the density related to a given excitation \cite{Esry.pra:98}. From the orbitals' response, Eq.~\eqref{eq:response}, we can calculate the time-dependent density for a probe frequency $\omega$ and probing fields $f^{\pm}$: 
\be\label{eq:density_response}
\rho(\mathbf{r},t)=\sum_{i=1}^M n_i|\phi_i(\mathbf{r},t)|^2\approx \sum_{i=1}^M n_i|\phi_i^0(\mathbf{r})|^2+2\sum_k\frac{\gamma_k}{\omega-\omega_k}\Delta\rho^k(\mathbf{r})\cos{(\omega t)}\,.
\ee
The density shows the largest response at the linear response resonance frequencies. For simplicity, we neglect here the typically very small imaginary parts of the response amplitudes, and assume real stationary orbitals $\phi_i^0(\mathbf{r})$.  We then obtain for the oscillatory part of the real space density
\be\label{eq:density_osc}
\Delta\rho^{k}(\mathbf{r})=\sum_{i=1}^M \sqrt{n_i}\phi_i^0(\mathbf{r})\left\{ u_i^k(\mathbf{r}) +v_i^{k}(\mathbf{r}) \right\}\,.
\ee

The density in momentum space provides information about coherence properties of the system. For example, when two initially spatially separated parts of a BEC interfere in time-of-flight experiments, the density in momentum space describes approximately the interference pattern which is obtained on average. For a coherent BEC the fringe contrast is high, whereas it is zero for a two-fold fragmented BEC \cite{pitaevskii:01}. We note that in general interactions have to be taken into account during  expansion and interference of the BECs, leading to interference fringes even for two independent BECs \cite{cederbaum.prl:07}. The Fourier transformed orbitals and amplitudes, which we denote by $\tilde{\phi}_i^0(\mathbf{p})$, $\tilde u_i^k(\mathbf{p})$ and $\tilde v_i^k(\mathbf{p})$, respectively, are in general complex. Therefore, the density oscillates at fixed amplitude but with a momentum dependent phase shift $\alpha^k(\mathbf{p})$:
\bea\label{eq:density_response_p}
&&\tilde{\rho}(\mathbf{p},t)=\sum_{i=1}^M n_i|\tilde{\phi}_i(\mathbf{p},t)|^2\approx \sum_{i=1}^M n_i|\tilde{\phi}_i^0(\mathbf{p})|^2+2\sum_k\frac{\gamma_k}{\omega-\omega_k}|\Delta\tilde{\rho}^k(\mathbf{p})|\cos{\left[\omega t-\alpha^k(\mathbf{p}) \right]}\,,\non
&&\alpha^k(\mathbf{p})=\arctan{\left\{\Im{\Delta\tilde\rho^k(\mathbf{p})}/\Re{\Delta\tilde\rho^k(\mathbf{p})}\right\}}\,.
\eea
The momentum-space density oscillations are given by 
\be\label{eq:density_osc_p}
\Delta\tilde\rho^k(\mathbf{p})=\sum_{i=1}^M \sqrt{n_i}\tilde{\phi}_i^{0,*}(\mathbf{p})\left[\tilde u_i^k(\mathbf{p})+\tilde v_i^{k}(\mathbf{p})\right]\,.
\ee
Note that in Eq.~\eqref{eq:density_response_p}  the modulus of the momentum-space density oscillations, $|\Delta\tilde\rho^k(\mathbf{p})|$, appears. This modulus  can be measured in experiments as the \emph{maximal} value of the density at each momentum $\mathbf{p}$. 

The density response in position space can for some special cases be directly connected to the response weights. For a real periodic driving, which can be  translated to $f(\mathbf{r})=f^+(\mathbf{r})=f^-(\mathbf{r})$, we can write Eq.~\eqref{eq:weight} alternatively as
\be\label{eq:weight_a}
\gamma_k=\int d\mathbf{r}f(\mathbf{r})\Delta\rho^{k}(\mathbf{r})\,.
\ee
For this case the response of any observable is proportional to the density oscillations.

\section{Application to Bose-Einstein condensates in double-well potentials \label{sec:app}}

Before presenting our linear response studies of BECs in one-dimensional symmetric and asymmetric double-well potentials, we will briefly discuss the structure of the (possibly fragmented) mean-field states in double-well potentials which are lowest in energy. 

\begin{figure}[h]
           \includegraphics[width=.95\columnwidth]{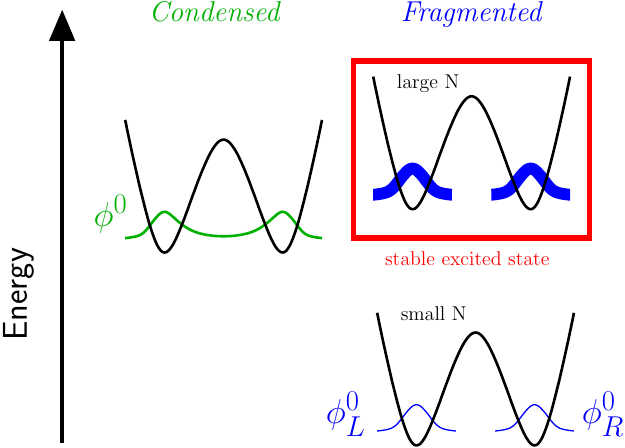}
  \caption{(Color online) Schematic comparison between condensed, fragmented and stable excited fragmented states in a double-well potential. The single orbital of a simple BEC [$\phi^0(x)$] is \emph{delocalized} (left chart). For a fragmented BEC the orbitals corresponding to different fragments [$\phi^0_L(x)$ and $\phi^0_R(x)$] are \emph{localized}   (right charts).  For a given barrier height, the stable fragmented state (lower right chart) is lower in energy than the condensed one up to some (possibly very high) atom number. Then, the condensed state becomes the lowest in energy. However, an energetically close stable excited fragmented state (upper right chart) typically exists. It is separated by an energy barrier from the condensed one. 
 \label{fig:meta}}
\end{figure}

Within a mean-field treatment, related to the ansatz of Eq.~\eqref{eq:MFansatz}, the ground state in such a trap is either condensed or two-fold fragmented, depending on the barrier height and the interaction strength \cite{streltsov.pra:05}. The many-body wave function for the condensed state reads 
\be\label{eq:MFansatzGP}
\Psi(x_1,...,x_N)=\Pi_{i=1}^N\phi^0(x_i)\,,
\ee
whereas for the two-fold fragmented state with degree of fragmentation $n/N$ it is given by
\be\label{eq:MFansatzBMF}
\Psi(x_1,...,x_N)=\hat{\mathcal S}\Pi_{i=1}^n\phi_L^0(x_i)\Pi_{j=n+1}^{N}\phi_R^0(x_j)\,.
\ee
Here, orbital $\phi_L^0(x)$ [$\phi_R^0(x)$] is localized in the left [right] well. The energies of those states are given by
\be\label{eq:en_gp}
E^{M=1}=N\int dx \phi^{0,*}(x)\hat h(x)\phi^0(x)+\frac{\lambda_0N(N-1)}{2}\int dx |\phi^0(x)|^4\,,
\ee
and
\bea\label{eq:en_bmf}
E^{M=2}&=&\sum_{i=L,R} \left[n_i\int dx \phi^{0,*}_i(x)\hat h(x)\phi^0_i(x)+\frac{\lambda_0 n_i(n_i-1)}{2}\int dx |\phi^0_i(x)|^4\right]\non
&+&2\lambda_0 n_L n_R\int |\phi_L^0(x)|^2|\phi_R^0(x)|^2 dx\,,
\eea
respectively. Above a critical barrier height, a fragmented state [Eq.~\eqref{eq:MFansatzBMF} with $n\neq 0$] becomes favorable in energy over a condensed one, Eq.~\eqref{eq:MFansatzGP}. The same thing happens when the inter-particle interaction strength $\lambda_0$ exceeds a critical value. Typically, these transition points shift with atom number N (at fixed $\lambda_0 N$) to higher barrier and/or stronger interaction strengths. Importantly, even when the condensed state is lower in energy than the fragmented one, above a critical interaction strength the latter can be considered a stable excited state \cite{streltsov1:04,streltsov.pra:05}. It is typically slightly higher in energy than the condensed state, and is separated from it by an energy barrier. For example, in a symmetric double-well both the condensed and 50-50 left-right fragmented states are local minima with respect to a change in the critical occupation. In Fig.~\ref{fig:meta} we depict schematically these states and their energies for symmetric double-wells.

In principle, besides the orbitals' excitations described by LR-BMF, there are excitations consisting of the redistribution of atoms between the orbitals (`hopping excitations'). Such processes can  most easily be described within a two-site Bose-Hubbard (BH) model \cite{jaksch:98}. In Appendix~\ref{app:lr_bh} we derive the linear response of Bose-Hubbard (`LR-BH'), i.e., the response to a time-dependent potential as in Eq.~\eqref{eq:pert}. The excitation energies coincide as can be expected with the eigenenergies of the BH model. 
We study the LR-BH response weights and see that the dominant physical processes in a double-well potential with a time-oscillating potential perturbation are the orbitals' (or spatial) excitations.

We note that the BMF and TDMF are mean field methods, and thus generally offer qualitative descriptions of BECs in double-well potentials \cite{alon.prl1:05,alon.prl2:05,cederbaum.prl:07}.  In order to capture effects beyond mean field, one has to employ a description where the ground state of BECs in a double-well potential is neither completely condensed nor two-fold fragmented. In this case the off-diagonal elements of the one-body reduced density operator (in the left--right basis) starts to play a crucial role \cite{sakmann:08}. The \emph{multiconfigurational Hartree for Bosons} (MCHB) method \cite{streltsov:06} and its time-dependent variant, the \emph{multiconfigurational time-dependent Hartree for Bosons} (MCTDHB) method \cite{streltsov:07,alon:08}, offer full many-body descriptions.  However, those methods are numerically much more demanding and are thus generally restricted to systems with smaller atom numbers and/or weaker interactions.

In the following we study ultra-cold bosons in a one-dimensional double-well potential parametrized as follows:
\be\label{eq:pot}
V(x)=b/2\cdot\cos{(\frac{\pi}{3}x)}+\omega_{ho}^2 x^2/2+a\cdot x\,,
\ee
with the barrier height $b$, harmonic oscillator frequency $\omega_{ho}$ determining the overall harmonic confinement, and asymmetry $a$. We present in the following excitation spectra of fragmented states, which originate from the derived LR-BMF response matrix Eq.~\eqref{eq:lrm} for the special case $M=2$ (the linear response matrix is given in Appendix~\ref{app:spec}).  To this we compare  the response of condensed states, obtained from the number-conserving version of the BdG equations, i.e., LR-GP (those equations are discussed in detail in Appendix~\ref{app:spec}). Throughout   this work we choose an harmonic confinement with $\omega_{ho}=\sqrt{2}$. We discuss different values of the barrier height $b$, as well as interaction strengths $\lambda_0 N$. 

The TDMF equations [see Eq.~\eqref{eq:TDMF}] are for large enough atom numbers (say, $N>20$) practically independent of the total atom number $N$, as long as $\lambda_0 N$ is kept fixed. They rather depend solely on the relative occupations $n_i/N$. Similar statements hold also for the linear response. In this work we use $N=100$ throughout, but we stress that the excitation spectra and corresponding observables are almost the same for larger atom numbers.

Our linear response studies are based on a small perturbation of ground states. The ground state orbital for a simple BEC is obtained from the stationary GP equation
\be\label{eq:stat_GP}
\hat H_{GP}\phi^0(x)=\mu\phi^0(x)\,.
\ee
For two-fold fragmented BECs the ground-state orbitals are calculated as the lowest in energy solution of the $M=2$ best-mean field equations \cite{cederbaum:03}:
\bea\label{eq:stat_BMF}
\left\{\hat h(x)+\lambda_0(n_L-1)|\phi_L^0(x)|^2+2\lambda_0 n_R|\phi_R^0(x)|^2\right\}\phi_L^0(x)=\mu_{LL}^0\phi_L^0(x)+\mu_{LR}^0\phi_R^0(x)\,,\non
\left\{\hat h(x)+\lambda_0(n_R-1)|\phi_R^0(x)|^2+2\lambda_0 n_L|\phi_L^0(x)|^2\right\}\phi_R^0(x)=\mu_{RR}^0\phi_R^0(x)+\mu_{RL}^0\phi_L^0(x)\,.
\eea
Those ground-state orbitals $\phi^0(x)$, $\phi_L^0(x)$ and $\phi_R^0(x)$ are real.

Technically, we determine the ground states by imaginary time propagation of Eqs.~\eqref{eq:stat_GP} and \eqref{eq:stat_BMF} until the energy has converged to the desired accuracy of $10^{-14}$. We use per orbital a grid size of $N_g=251$ points and a box of size $12$. The kinetic part is solved utilizing Fast Fourier transform. Using more grid points and/or a larger box does not lead to any visible differences in our plots. Then, to determine the frequencies $\omega_k$ we diagonalize the non-Hermitian linear response matrix Eq.~\eqref{eq:lrm} on the same grid (per response amplitude).  
 We  concentrate on the positive branch of eigenvalues $\omega^k$ (we recall that there is a negative partner to each eigenvalue, see Section~\ref{sec:der}).

\subsection{Symmetric double-well \label{subsec:sym}}

We start with a symmetric double-well potential, which is given by Eq.~\eqref{eq:pot} with zero asymmetry $a=0$.  We first discuss the dependence of the excitations on the interaction strength and choose a high barrier in Sec.~\ref{subsubsec:high}. Even for weak interaction strengths, the response of LR-GP and LR-BMF in momentum space is qualitatively different. For larger interactions two types of excitations emerge within LR-BMF and become energetically separate.  Thereafter we proceed to study how the response changes with barrier height in Sec.~\ref{subsubsec:low}.

\subsubsection{High barrier \label{subsubsec:high}}

\begin{figure}[h]
           \includegraphics[width=.95\columnwidth]{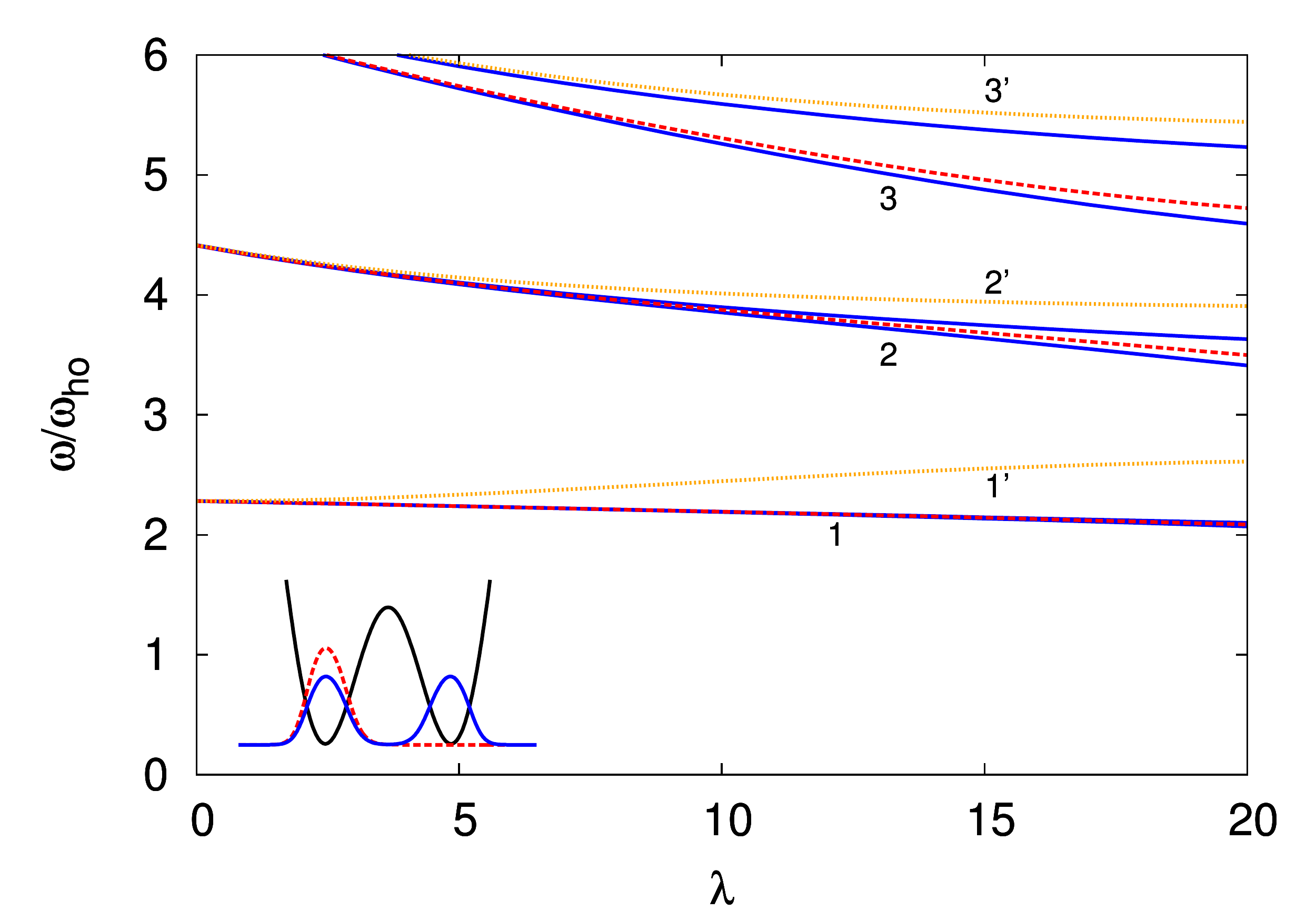}
  \caption{(Color online) Excitation spectra of a BEC in a symmetric double-well potential ($a=0$) versus interaction strength with high barrier height ($b=20$). The total atom number is $N=100$, and the overall harmonic confinement given by $\omega_{ho}=\sqrt{2}$.  We compare the linear response of LR-GP, shown by the blue solid lines, and LR-BMF, shown by the red dashed lines for direct, and orange dotted lines for swapped excitations. The excitations are grouped into pairs of lines with gerade-ungerade symmetry and marked with numbers. The swapped excitations of LR-BMF are marked with primed numbers. Inset: We plot the potential by the black  solid line. The corresponding ground state orbital of GP for $\lambda_0 N=10$ is shown by the blue solid,  and the left orbital of BMF by the red dashed line. All quantities are dimensionless. 
 \label{fig:spectrumL20}}
\end{figure}

For high barrier heights the orbitals practically vanish around $x=0$ as can be seen in the inset of Fig.~\ref{fig:spectrumL20}. 
As a consequence, the left (right) orbital of BMF has a shape similar to the left (right) half of the GP equation. Moreover, all terms in the LR-BMF response matrix, Eq.~\eqref{eq:lrm}, which are proportional to $|\phi_R^0(x)|\cdot|\phi_L^0(x)|$, are very small. Hence, the eigenvalue problem of Eq.~\eqref{eq:evL} can be written as
\bea\label{eq:uL}
&&\hat P\left[\left(\hat Z^{0,'}-\mu_{LL}^0\right) u_L^k-\mu_{LR}^0 u_R^k + \tilde n\left(\phi_L^0\right)^2 v_L^k\right]=\omega u_L^k\,,\non
&&\hat P\left[\left(\hat Z^{0,'}-\mu_{LL}^{0,*}\right) v_L^k-\mu_{LR}^{0,*} v_R^k + \tilde n\left(\phi_L^{0,*}\right)^2 u_L^k\right]=-\omega v_L^k\,.
\eea
Similar equations hold for $u_R^k$ and $v_R^k$ (with indices $L$ and $R$ interchanged), 
see Appendix~\ref{app:spec} for the full matrix. We defined here $\tilde n=\lambda_0N/2$, and approximated $N\approx N-1$.
Most importantly, $u_L^k$ and $v_L^k$ are governed by the same operator $\hat Z^{0,'}$ as $u_R^k$ and $v_R^k$.
\be
\hat Z^{0,'}=\hat h +2\tilde{n}\left(|\phi_L^0|^2+|\phi_R^0|^2\right)\,.
\ee
 Note the difference of $\hat Z^{0,'}$ to the TDMF-operators $\hat Z_i^0$, which carry the index $L$ or $R$. Hence, the first term in each line of Eq.~\eqref{eq:uL} describes a particle in the effective potential of two condensates, both carrying a factor of two due to exchange interactions \cite{alon.pla:07}. 
 The term proportional to the off-diagonal Lagrange multipliers $\mu_{LR}^0$  and $\mu_{RL}^0$ is a coupling term to the other amplitude $u_R^k$. The last terms on the left hand sides couple $u_L^k$ to $v_L^k$. 

For weak interaction strengths, the effects of coupling of $u_L^k$ to $u_R^k$ and $v_L^k$ (and similarly for $L\leftrightarrow R$) are negligible. Since  
$\hat Z^{0,'}$ is symmetric in $L$ and $R$, it originates to delocalized response amplitudes which have either gerade or ungerade symmetry.\footnote{We note that for smaller atom numbers, on the order of $N=100$, the difference between $N$ and $N-1$ leads to localized orbitals. However, in this regime the two basis sets (i.e., left-right or gerade-ungerade) lead to the same physics.} LR-GP reduces to an equation similar to that for $u_L^k$ or $u_R^k$ of LR-BMF. Hence, the energies [see Fig.~\ref{fig:spectrumL20}] and response amplitudes coincide. However, as we will show later, the momentum space density responses  of LR-GP and LR-BMF differ strongly due to the different structures of the ground states, i.e., coherent or fragmented [see Eqs.~\eqref{eq:MFansatzGP} and \eqref{eq:MFansatzBMF}, respectively].  



\begin{figure}[h]
           \includegraphics[width=.99\columnwidth]{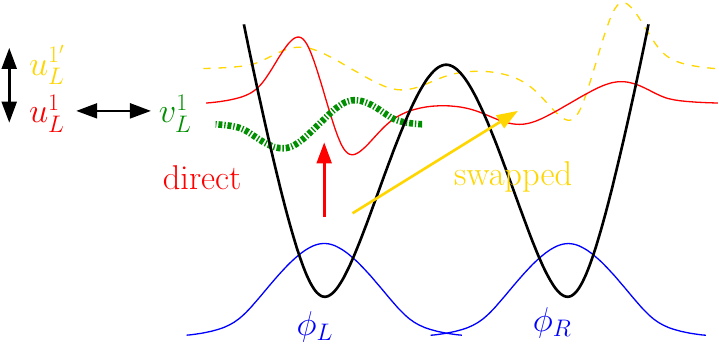}
  \caption{(Color online) Sketch of the two types of excitations occurring for fragmented BECs in a double-well potential. Direct (swapped) excitations can be interpreted as the promotion of a boson from one orbital to an excited state which dominates in the same (other) well, e.g., from $\phi_L$ to $u_L^1$ ($u_L^{1'}$). The transfer of atoms to direct (red solid line) costs less energy than to swapped (orange dashed lines) excitations. This is because a depletion, related to $v_L^1$ (green dashed-dotted line), reduces the energy of direct excitations. For the swapped excitations, $v_L^{1'}\approx 0$. Response weights of direct and swapped excitations of LR-BMF are proportional to the overlap between an orbital and the corresponding response amplitudes. Hopping excitations of LR-BH are proportional to the much smaller overlap of the two ground-state orbitals.
 \label{fig:exch}}
\end{figure}

For larger interaction strengths, we find that for LR-BMF two types of excitations become energetically separated. 
In particular, an excitation can be either to an orbital, which dominates in the same well (`direct' excitation), or in the other well (`swapped' excitation). 
We sketch the notion of swapped excitations in Fig.~\ref{fig:exch}. 
While the orbitals $\phi_L$ and $\phi_R$ are localized and have very small overlaps, the $u$-amplitudes of LR-BMF are partly delocalized. 

\begin{figure}[h]
          \includegraphics[width=.99\columnwidth]{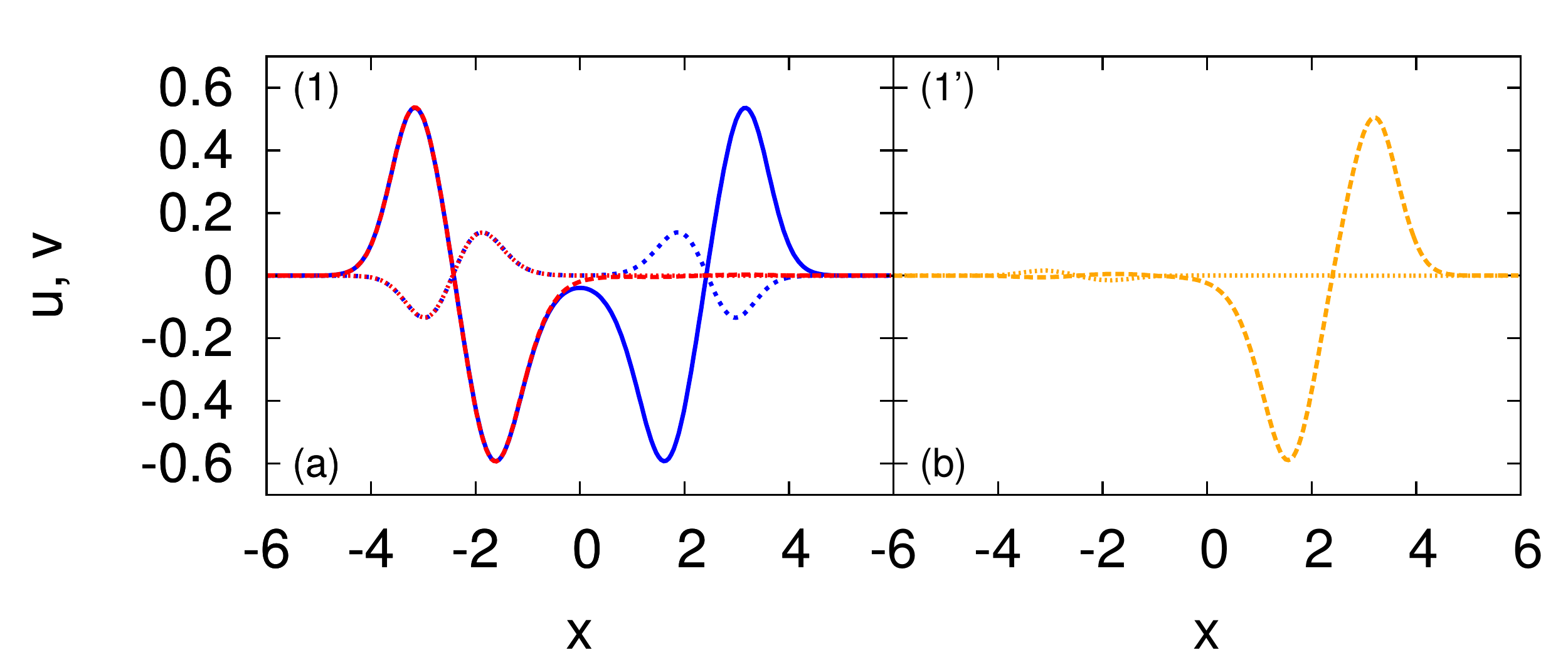}
  \caption{(Color online)  Example of response amplitudes $\mus^k$ and $\mvs^k$ for the same double-well system as in Fig.~\ref{fig:spectrumL20} and for $\lambda_0 N=10$. (a) The amplitudes $u$ and $v$ of LR-GP, shown by the blue solid and short-dashed lines, respectively, are symmetric functions. For LR-BMF, instead, they are localized functions. We show $u_L^{1g}$ by the red dashed line, and  $v_L^{1g}$ by the red dotted line. (b) Amplitudes of the swapped excitations $u_L^{1g'}$ and $v_L^{1g'}$ of LR-BMF, shown by the orange dashed and dotted lines, respectively. Amplitudes of LR-BMF are magnified by a factor of $\sqrt{2}$ for easier comparison. See text for more details. All quantities are dimensionless.
 \label{fig:uv_ex}}
\end{figure}

We can understand the energetical splitting between direct and swapped excitations as follows. In Eq.~\eqref{eq:uL}, the term which accounts for a coupling to $v_L^k$  becomes important for larger interactions. The $v$-amplitudes describe a depletion of the true ground state of a condensate in which a few atoms occupy excited states \cite{esry:97}.  Furthermore, we find the $v$-amplitudes, and, hence, the depletion, to be local. They are nonzero only for direct excitations, see Fig.~\ref{fig:uv_ex}. The depletion thus lowers the energy of direct excitations as compared to swapped excitations where the energy is determined solely by exchange interactions [see red dashed and orange dotted lines in Fig.~\ref{fig:spectrumL20}, respectively]. We conclude that \emph{we found a class of excitations in a fragmented system which do not appear at all in a condensed one.}

Accordingly, for a fragmented state the amplitudes are localized, in contrast to LR-GP. The response amplitudes have either gerade (g) or ungerade (u) symmetry, which we label as $k=1g,1u$, where $k$ is the index of the excitation. We show the gerade ones $u^{1g}(x)$ and $v^{1g}(x)$  in Fig.~\ref{fig:uv_ex} (a) by the blue solid and dotted lines, respectively. 
The linear combination of left and right LR-BMF amplitudes is either gerade or ungerade, and we thus use the same labeling as for the condensed state.  We show in Fig.~\ref{fig:uv_ex} (a) the first direct excitation of LR-BMF. In particular we plot the LR-BMF amplitudes $u_L^{1g}(x)$ and $v_L^{1g}(x)$, which have for $x<0$ the same shape as for LR-GP (except around $x=0$).   We further note that the response amplitudes in both condensed and fragmented cases have a node which ensures orthogonality with respect to the ground-state orbitals. In Fig.~\ref{fig:uv_ex} (b) we show the first swapped excitation of LR-BMF, i.e., the amplitudes $u_L^{1'g}(x)$ and $v_L^{1'g}(x)$. The response amplitude $v_L^{1'g}(x)$ almost vanishes.

The terms of Eqs.~\eqref{eq:uL} proportional to the Lagrange multipliers $\mu^0_{LR}$ and $\mu^0_{RL}$, as well as the  terms proportional to the orbitals as $\phi_L^0\cdot\phi_R^0$, which we neglected in Eq.~\eqref{eq:uL}, induce another small shifts of energies.


The response weights of swapped excitations are given by the overlap integrals of orbitals, response amplitudes, and perturbations: 
\be
\gamma_{1'}=\sqrt{N/2}\int dx\left(u_L^{1',*}f^+\phi_L^0+u_R^{1',*}f^+\phi_R^0+v_L^{1',*}f^-\phi_L^0+v_R^{1',*}f^-\phi_R^0\right)\,.
\ee
Although it comprises a transfer of atoms to the other well, it is in general dominant over excitations involving a  hopping of atoms between the orbitals. Within linear response of the Bose-Hubbard model (`LR-BH', for details we refer to Appendix~\ref{app:lr_bh}) and under a periodic potential perturbation, hopping can occur either directly through tunneling, or mediated by a time-dependent potential difference between the wells. The first process is completely irrelevant here since the orbitals $\phi_L^0(x)$ and $\phi_R^0(x)$ have small overlaps. The second process has finite response weights only for the first few hopping excitations, with energies well below the first excitation of LR-BMF. 
 

We move on to the study of excitations by means of an observable. Let us first discuss the limit of very weak interaction strengths by means of approximate formulas for the orbitals and response amplitudes. In particular, we denote with $\psi^n(x)$ ($n=0,1,2,...$) the normalized ground and $n$-th excited harmonic oscillator eigenfunction, centered around $x=0$. We model the GP orbital and first excited $u$-amplitudes of LR-GP as (the $v$-amplitudes are negligible for weak interaction strengths):
\bea\label{eq:simple_GP}
&&\phi^0(x)=\left[\psi^0(x_L)+\psi^0(x_R)\right]/\sqrt{2}\,,\non
&& u^{1g,u}(x)=\left[\psi^1(x_L)\mp\psi^1(x_R)\right]/\sqrt{2}\,,
\eea
with $x_{L}:=x+d$ ($x_{R}:=x-d$). $d$ is half the distance between the minima of the left and the right wells. The minus in front of $\psi^1(x_R)$ is needed in order to construct a gerade function out of two displaced ungerade functions $\psi^1(x_{L})$ and $\psi^1(x_{R})$. The BMF orbitals and LR-BMF response amplitudes can be considered to be completely localized. For direct excitations we have
\bea\label{eq:simple_BMF}
&&\phi^0_L(x)=\psi^0(x_L)\,,\quad\phi^0_R(x)=\psi^0(x_R)\,,\non
&& u_L^{1g,u}(x)=\psi^1(x_L)/\sqrt{2}\,,\quad u_R^{1g,u}(x)=\mp\psi^1(x_R)/\sqrt{2}\,.
\eea
The normalization of the $u$-amplitudes follows from the orthonormalization relations in Eq.~\eqref{eq:orth}.   The density oscillations of the direct excitations are obtained by plugging Eqs.~\eqref{eq:simple_GP} and \eqref{eq:simple_BMF} into Eq.~\eqref{eq:density_osc}:
\bea
&&\frac{\left[\Delta\rho^{1g,u}\right]_{GP}}{\sqrt{N}}=\left[\psi^0(x_L)+\psi^0(x_R)\right]\left[\psi^1(x_L)\mp\psi^1(x_R)\right]/2\,,\non
&&\frac{\left[\Delta\rho^{1g,u}\right]_{BMF}}{\sqrt{N}}=\left[\psi^0(x_L)\psi^1(x_L)\mp\psi^0(x_R)\psi^1(x_R)\right]/2\,.
\eea
  When assuming that the overlap between displaced functions vanishes, it results that $\left[\Delta\rho^{1g,u}\right]_{GP}=\left[\Delta\rho^{1g,u}\right]_{BMF}$. Similarly, for the response weights  holds $[\gamma_{1g,u}]_{GP}=[\gamma_{1g,u}]_{BMF}$. Moreover, the density response of the swapped excitations of LR-BMF  and their response weights vanish. Thus, for large barriers and weak interaction strengths, the density in position space responds   in exactly the same fashion for both condensed and fragmented states. 

But what if we proceed to momentum space? In this case, the ground state densities are \emph{qualitatively} different: whereas for GP the density shows up a modulation due to the coherence between the bosons in the left and right well, the density of a fragmented state is simply a Gaussian.  Using a similar notation as above, we denote the Fourier transformed ground and excited harmonic oscillator states as $\tilde\psi^n(p)$ ($n=0,1,2,...$). We remind that the Fourier transform of a translated function amounts to the Fourier transform of the original function times an additional phase factor, i.e., $FT\left[\psi_n(x-d)\right]=e^{-ipd}FT\left[\psi_n(x)\right]$. We then find for the condensed system as Fourier transforms of Eqs.~\eqref{eq:simple_GP}
\bea\label{eq:up_gp}
&&\tilde\phi^{0}(p)=\sqrt{2}\cos{(pd)}\tilde\psi^0(p)\,,\non
&&\tilde u^{1g}(p)=\sqrt{2}i\sin{(pd)}\tilde\psi^1(p)\,,\non
&&\tilde u^{1u}(p)=\sqrt{2}\cos{(pd)}\tilde\psi^1(p)\,,
\eea
and for the fragmented one as Fourier transforms of Eqs.~\eqref{eq:simple_BMF}
\bea\label{eq:up_bmf}
&&\tilde\phi^{0}_{L/R}(p)=\tilde\psi^0(p)e^{\pm ipx}\,,\non
&&\tilde u_{L}^{1g,u}(p)=\tilde\psi^1(p)e^{ipx}/\sqrt{2}\,,\non
&&\tilde u_{R}^{1g,u}(p)=\mp\tilde\psi^1(p)e^{- ipx}/\sqrt{2}\,.
\eea
Plugging Eqs.~\eqref{eq:up_gp} and \eqref{eq:up_bmf} into Eq.~\eqref{eq:density_osc_p}, we obtain for LR-GP the following density oscillations in momentum space 
\bea\label{eq:estp}
&&\frac{\left|[\Delta\tilde\rho^{1g}(p)]_{GP}\right|}{\sqrt{N}}=\tilde{\psi}^0(p)\left|\tilde{\psi}^1(p)\sin{(2pd)}\right|\,,\non
&&\frac{\left|[\Delta\tilde\rho^{1u}(p)]_{GP}\right|}{\sqrt{N}}=2\tilde{\psi}^0(p)\left|\tilde{\psi}^1(p)\right|[\cos{(pd)}]^2\,,
\eea
which show up modulations of the phase with frequencies $2pd$ and $pd$, respectively.
For the first direct excitation of LR-BMF, marked as 1, we have
\bea\label{eq:p_phase}
&&\left|[\Delta\tilde\rho^{1'g}(p)]_{BMF}\right|=0\,,\non
&&\frac{\left|[\Delta\tilde\rho^{1'u}(p)]_{BMF}\right|}{\sqrt{N}}=\tilde{\psi}^0(p)\left|\tilde{\psi}^1(p)\right|\,.
\eea
Thus, the momentum-space density response vanishes for the gerade direct excitation and the ungerade direct one has only one node.  Thus, \emph{even for a high barrier and weak interaction strengths, the momentum-space density oscillations of condensed and fragmented BECs are qualitatively different.}

For completeness we quote here also the momentum-space density response for the first pair of swapped excitations, marked as 1', although the  corresponding response weights vanish for weak interaction strengths. It can be obtained by switching the signs in the exponents of the second and third quantities of Eqs.~\eqref{eq:up_bmf}. Plugged into Eq.~\eqref{eq:density_osc_p} this results in
\bea\label{eq:p_phase_s}
&&\frac{\left|[\Delta\tilde\rho^{1g}(p)]_{BMF}\right|}{\sqrt{N}}=\tilde{\psi}^0(p)\left|\tilde{\psi}^1(p)\sin{(2pd)}\right|\,,\non
&&\frac{\left|[\Delta\tilde\rho^{1u}(p)]_{BMF}\right|}{\sqrt{N}}=\tilde{\psi}^0(p)\left|\tilde{\psi}^1(p)\cos{(2pd)}\right|\,.
\eea
We find that the gerade-type density response of LR-GP  and the swapped gerade one of LR-BMF are very similar and have the same period. This is in contrast to the ungerade-type response, where for the swapped excitations of LR-BMF the period  doubles. 

\begin{figure}[h]
           \includegraphics[width=.99\columnwidth]{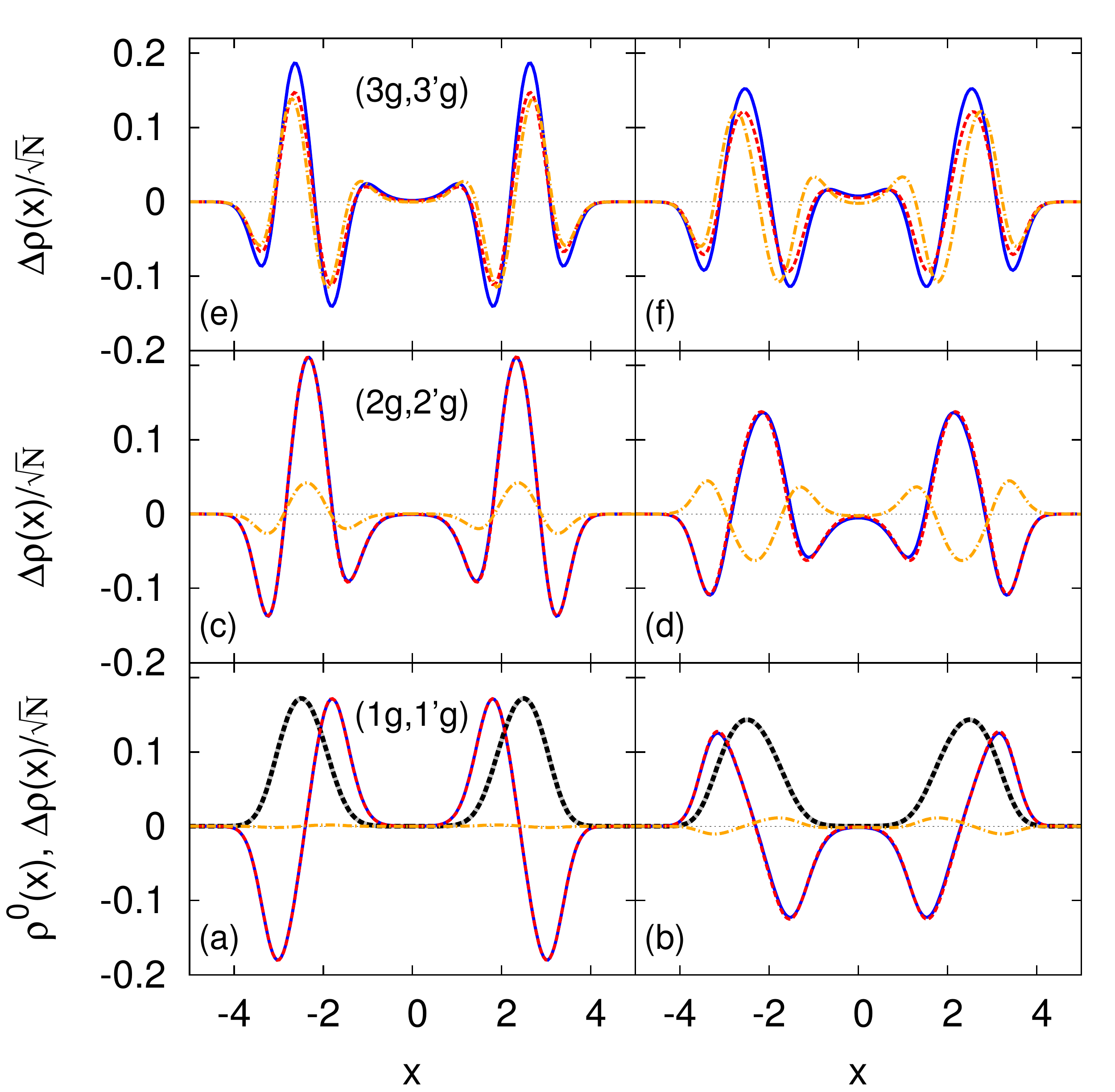}
  \caption{(Color online) Position-space densities and density oscillations for the same double-well system as in Fig.~\ref{fig:spectrumL20}. The interaction strength is $\lambda_0 N=10$ in the left panels, and $\lambda_0 N=20$ in the right panels. The ground state density of GP is shown in (a,b) by the broad gray solid line and is scaled for better comparison to the density of BMF plotted by the broad black dashed line. The GP and BMF results are seen to coincide. Gerade density oscillations of the indicated excited states are shown in (a-f). The LR-GP results are shown by the blue solid  lines. The dashed red lines (orange dashed-dotted lines) show results of LR-BMF for direct (swapped) excitations.  See text for more details. All quantities are dimensionless.
 \label{fig:dens20}}
\end{figure}

We now proceed to the density response at stronger interaction strengths. In Fig.~\ref{fig:dens20} we plot the density oscillations in position space as defined in Eq.~\eqref{eq:density_osc} for different excitations.
The broad gray solid and dark dashed lines in panels (a) and (b) show the ground state densities for GP and BMF, respectively. They perfectly coincide  for high barriers. In the left panels we show results for $\lambda_0 N=10$, and in the right ones for $\lambda_0 N=20$. It is interesting to note that the $v$-amplitudes have typically opposite signs as the $u$-amplitudes (see Fig.~\ref{fig:uv_ex}). 
They become important for large interaction strengths, and lead to a damping of the density oscillations  (compare left and rights panels in Fig.~\ref{fig:dens20}). For the lowest excitation, marked as 1 [see Fig.~\ref{fig:dens20} (a) and (b)], we find that the density oscillations of the condensed (blue solid lines) and fragmented states (red dashed lines) have almost the same shapes.  We plot only the gerade density oscillations, but there exist counterparts of ungerade type as well. Whenever gerade and ungerade excitations are energetically degenerate, the moduli of their real space density oscillations are on top of each other.\footnote{The response at a degenerate frequency is the sum of both the gerade and ungerade contributions, multiplied by the corresponding response weights Eq.~\eqref{eq:weight}.}  

For the swapped excitations of LR-BMF, the response amplitudes are partly delocalized [see Figs.~\ref{fig:exch} and \ref{fig:uv_ex}] and thus the position-space density response is non-negligible even for  the lowest swapped excitations at $\lambda=20$, marked as 1' [see Fig.~\ref{fig:dens20} (b)]. Similarly, the response weights are non-negligible.

\begin{figure}[h]
           \includegraphics[width=.99\columnwidth]{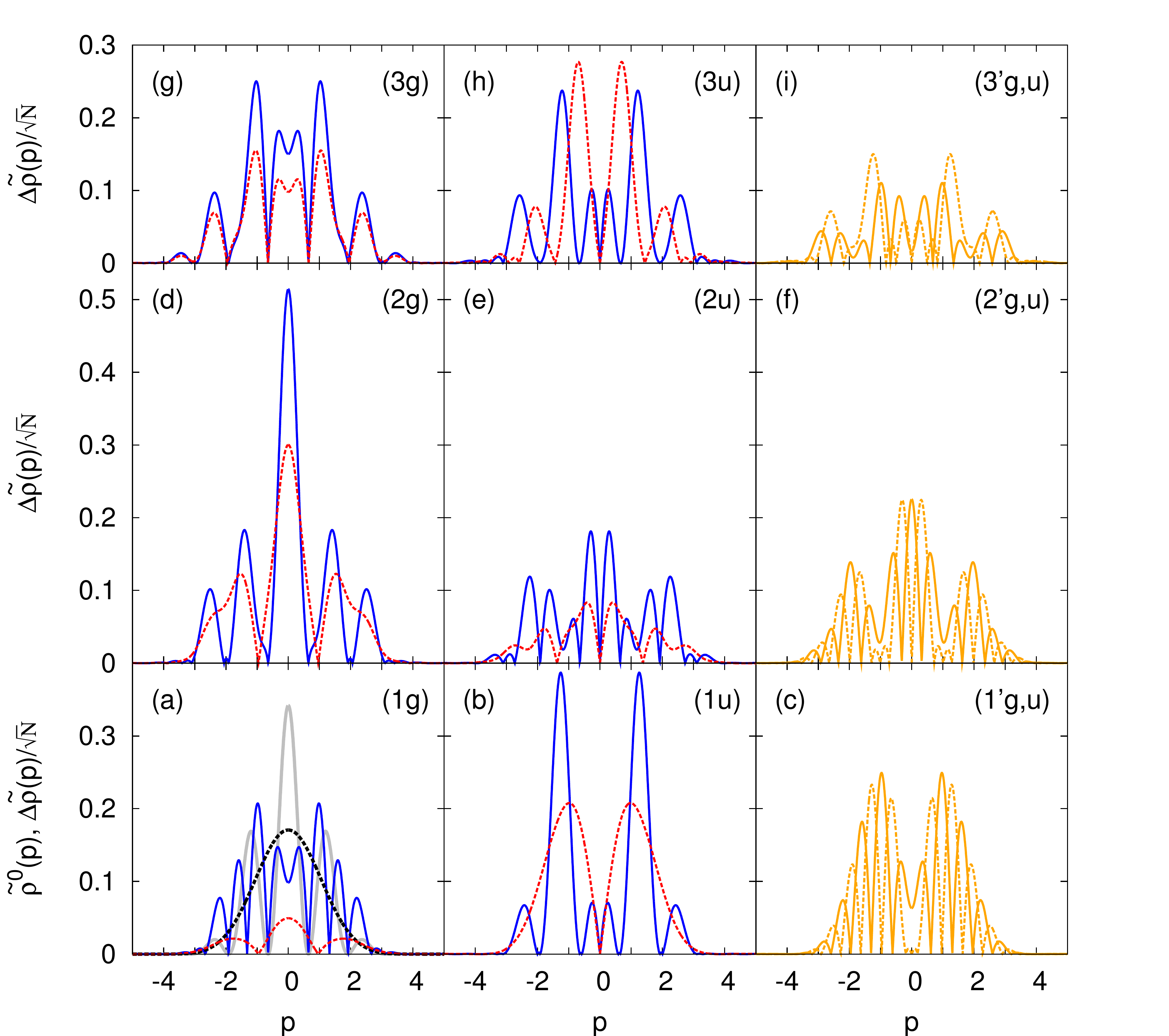}
  \caption{(Color online) Momentum-space densities and density oscillations for the same double-well system as in Fig.~\ref{fig:spectrumL20}. The interaction strength is $\lambda_0 N=10$. The ground state density of GP is shown in (a) by the broad gray solid line and is scaled for better comparison to the ground state density of BMF, which is shown by the broad black dashed  line. In momentum space they are completely different. Momentum-space density oscillations of gerade type of the indicated excited states (see Fig.~\ref{fig:dens20}) are shown in (a,d,g), and those of ungerade type in (b,e,h). Blue solid lines correspond to density oscillations of LR-GP and  red dashed lines to direct excitations of LR-BMF. Swapped excitations of LR-BMF are shown in (c,f,i). Orange solid (dashed) lines correspond to gerade-type (ungerade-type) density oscillations.  See text for more details. All quantities are dimensionless.
 \label{fig:densp20}}
\end{figure}

In momentum space, the GP ground state, plotted by the gray line in Fig.~\ref{fig:densp20} (a), shows an interference pattern reflecting the coherence between left and right condensates. 
This is completely different for BMF, plotted by the dashed black line, which describes two independent condensates. 
The momentum-space density response of LR-GP and LR-BMF for the excitations marked as 1 and 1' [see Fig.~\ref{fig:densp20} (a,b,c)] are at larger interactions still qualitatively described very well by Eqs.~\eqref{eq:estp}, \eqref{eq:p_phase_s} and \eqref{eq:p_phase} (note that the behavior around $p=0$ is not captured well by the simple equations for the orbitals and amplitudes, and the gerade solutions do not vanish at this point).

In the next higher group of excitations, marked as 2 and 2', also the direct excitations of LR-BMF deviate from the LR-GP energies, as can be seen in Fig.~\ref{fig:spectrumL20}. 
For LR-GP we find a splitting between gerade and ungerade excitations. The excitation energies of LR-BMF lie between them. The position space density response of the swapped excitations becomes more sizable as for the lowest excitation, see Figs.~\ref{fig:dens20} (d,e,f). Thus, swapped excitations become important for excitations with energies of the order of the barrier height. Whenever the swapped excitations give rise to an appreciable position-space response, this signifies the \emph{transition from below the barrier to above the barrier excitations} for a fragmented system, inasmuch as the lifting of the g-u degeneracy does for the response of a condensed system.



For the excitations marked as 3 and 3', the amplitudes become now more and more delocalized. This is a consequence of the fact that the higher the energy of the excitation, the less important is the barrier. While becoming similar in size, the direct excitations dominate in their own well, and the swapped ones in the respective other well. As a consequence,  within BMF we find a position-space density response weaker by a factor of $\sqrt{2}$ as compared to LR-GP, see Fig.~\ref{fig:dens20} (e,f). For weak interaction strengths this can be understood replacing in Eqs.~\eqref{eq:simple_BMF} 
\be
u_{L}^{1g,u}(x)=u_{R}^{1g,u}(x)=\left[\psi^3(x_L)\mp\psi^3(x_R)\right]/2\,.
\ee
From this we obtain $\left[\Delta\rho^{1g,u}\right]_{BMF}=\left[\Delta\rho^{1g,u}\right]_{GP}/\sqrt{2}$, as well as $[\gamma_{1g,u}]_{BMF}=[\gamma_{1g,u}]_{GP}/\sqrt{2}$. Qualitatively, these formulas describe also the density response for stronger interaction strengths.

\begin{figure}[h]
           \includegraphics[width=.99\columnwidth]{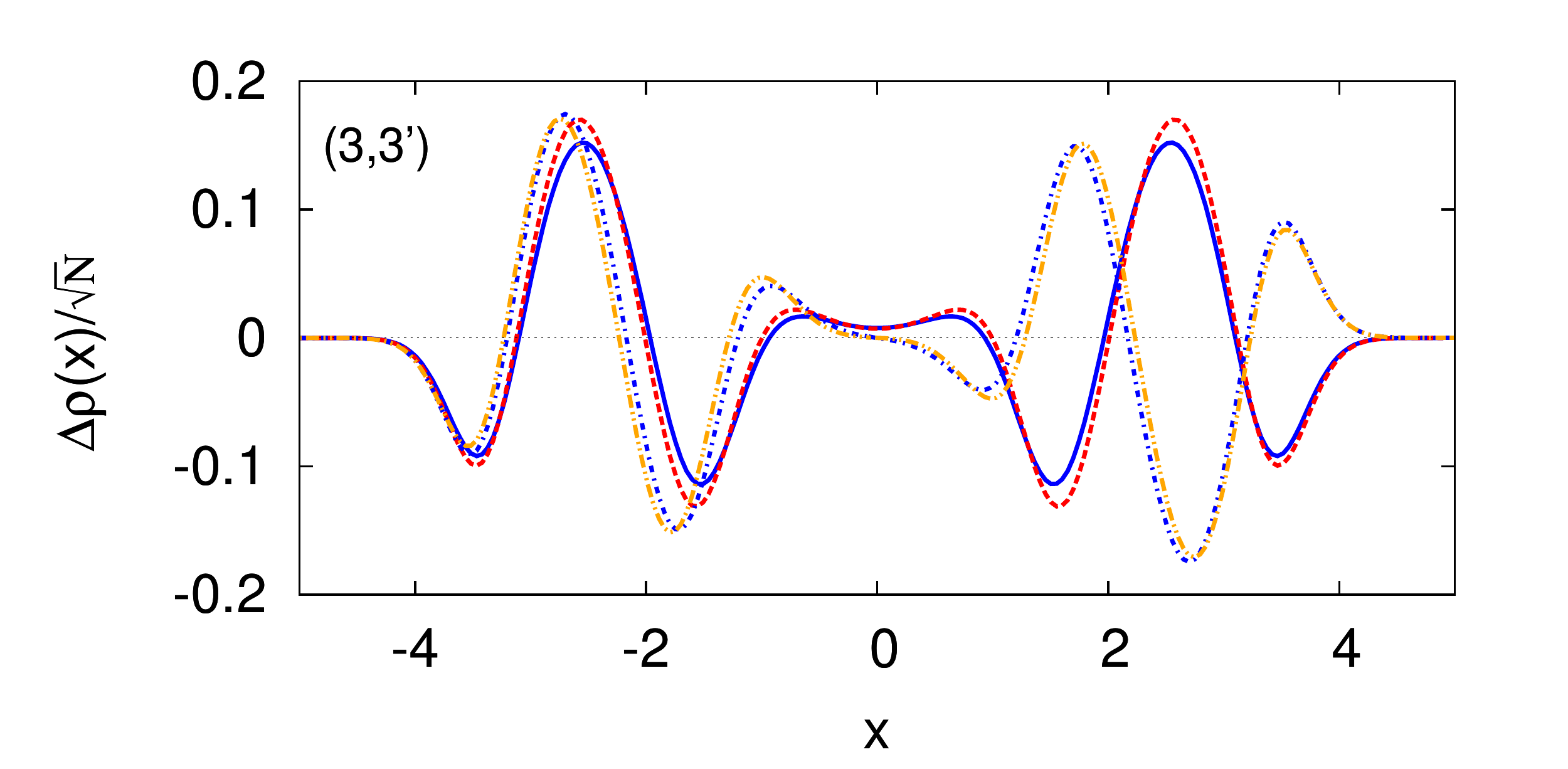}
  \caption{(Color online) Example of position-space density oscillations for the same double-well system as in Fig.~\ref{fig:spectrumL20}
   and for $\lambda_0 N=20$. Gerade [see Fig.~\ref{fig:dens20} (f)] as well as ungerade density oscillations of LR-GP are shown by the blue solid and dashed-dotted  lines, respectively. Direct gerade (swapped ungerade) density oscillations of LR-BMF are shown by the red dashed (orange dashed-double dotted) lines, and are multiplied by a factor of $\sqrt{2}$ for easier comparison. See text for more details. All quantities are dimensionless.
 \label{fig:dens_ex}}
\end{figure}

To gain further insight, we plot in Fig.~\ref{fig:dens_ex} both gerade and ungerade position-space density oscillations for LR-GP  with $\lambda_0 N=20$, as shown by the blue solid and dashed-dotted lines, respectively. 
 The shapes of the gerade and the ungerade excitations are different, reflecting the lifted energetical  degeneracy of the excitations. For the fragmented system we have two pairs of degenerate excitations: direct and swapped ones. We plot the gerade direct density response by the red dashed line, and the ungerade swapped one by the orange dashed-double dotted line. For a simpler comparison between LR-GP and LR-BMF we compensate for the factor of $\sqrt{2}$.
We conclude that well above the barrier the direct excitations of LR-BMF become similar to the gerade ones of LR-GP, although they have different energies [see Fig.~\ref{fig:spectrumL20}], and the swapped excitations become similar to the ungerade ones. Similar statements hold in momentum space, see Fig.~\ref{fig:densp20} (g,h,i).

For LR-GP there is in principle also an excitation where an atom is transferred to the ungerade solution of the GP equation. However, for the case of a high barrier the gerade and ungerade solutions of GP are almost degenerate, even for large interactions. Therefore this excitation has very small energy for all values of $\lambda$. Since it cannot be distinguished from zero in the plot, we do not show it in Fig.~\ref{fig:spectrumL20}. For the linear response of a BEC prepared in the ungerade GP state, the parity of the position space density oscillations changes. For the momentum space density oscillations, the periods of gerade and ungerade would be reversed as compared to Eqs.~\eqref{eq:estp}.

\subsubsection{Low barrier \label{subsubsec:low}}
\begin{figure}[h]
           \includegraphics[width=.95\columnwidth]{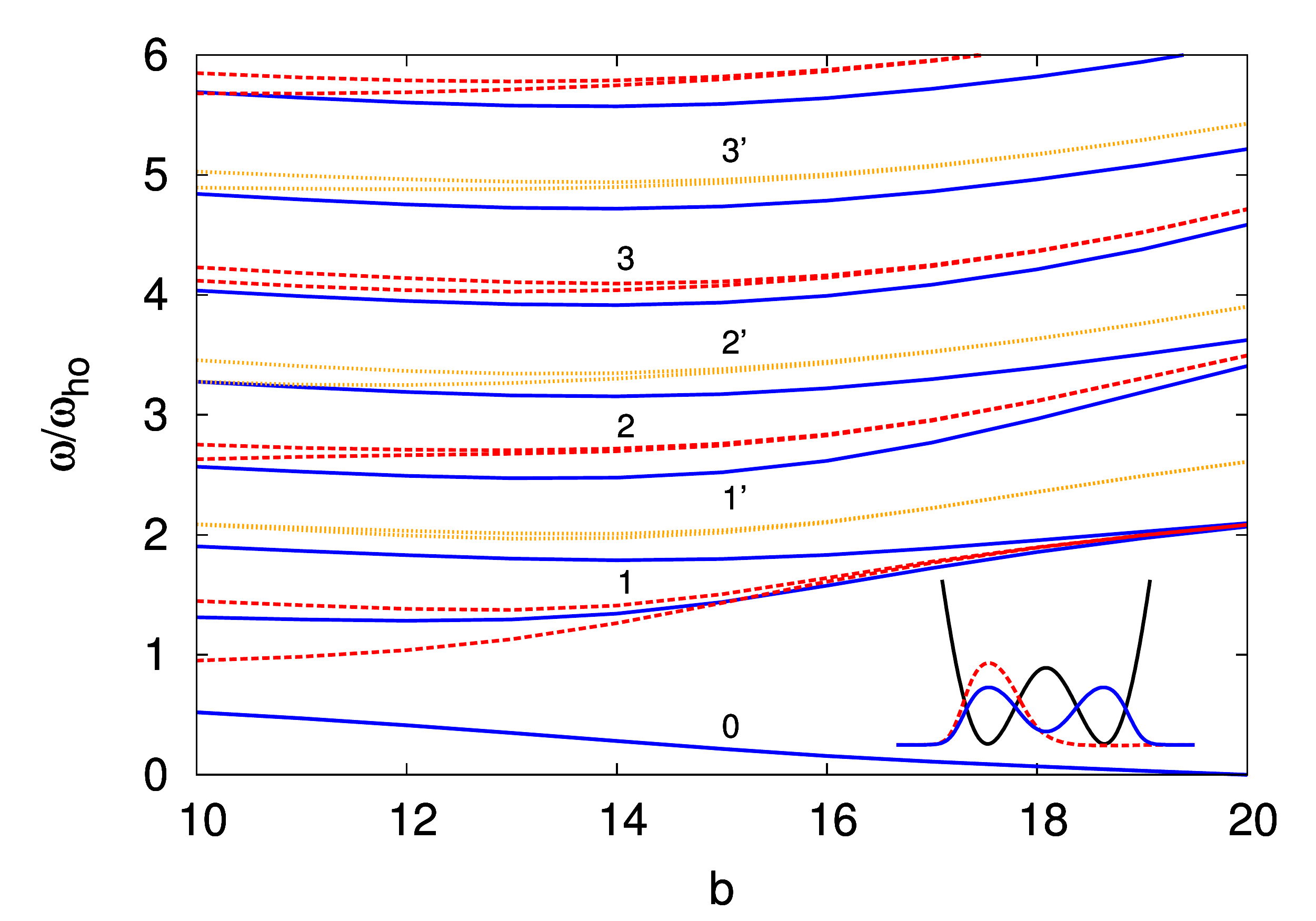}  
  \caption{(Color online)  Excitation spectra of a BEC in a symmetric double-well potential ($a=0$) versus barrier height for a fixed interaction strength $\lambda_0 N=20$. The total atom number is $N=100$, and the overall harmonic confinement given by $\omega_{ho}=\sqrt{2}$. We compare the linear response of LR-GP, shown by the blue solid lines, and LR-BMF, shown by the red dashed lines for direct, and orange dotted lines for swapped excitations. The excitations are grouped into pairs of lines with gerade-ungerade symmetry and marked with numbers. The swapped excitations of LR-BMF are marked with primed numbers. Inset:  We plot the potential by the black  solid line. The corresponding ground state orbital of GP for $b=14$ is shown by the blue solid,  and the left orbital of BMF by the red dashed line. All quantities are dimensionless.
 \label{fig:spectrumB}}
\end{figure}

with it 
Now we study what happens if we lower the barrier. In this case the orbitals of course penetrate the barrier more than at high barriers, see the inset of Fig.~\ref{fig:spectrumB}. 
Here we observe an excitation of LR-GP which lies energetically below all other excitations (labeled 0). This excitation corresponds to the ungerade solution of GP. When decreasing the barrier height, the energy of excitation 0 increases. This signifies, that for low barriers the ground state is condensed.

However, also fragmented states in a trap with low barrier could be of relevance in experiments. For example,  when one cools down a thermal gas in double-well potential, it is not clear whether one manages to condense the atoms really into the ground state of the system, or the system resides in stable fragmented excited states \cite{hofferberth:08}. Moreover, when two initially independent condensates are slowly merged, the question arises, whether the system evolves adiabatically and becomes condensed, or stays fragmented. When analyzing excitation spectra of a fragmented BEC using LR-BMF, the excitation 0 is absent, because there is no phase relation between the left and right condensates. Thus, the existence of an excitation to an ungerade mode can be used as a signature for coherence in the system and to characterize the state obtained by cooling into a double-well potential. 

For the higher excitations of LR-GP, we clearly see how they become degenerate as the barrier is raised. For example, around $b\approx 18$ the lowest pair of excitations, marked as 1, becomes degenerate (blue solid lines). For a fragmented ground state the situation is different: At small barriers (say $b\lesssim 15$), LR-BMF has twice as many non-degenerate excitations as compared to LR-GP. This splitting is due to the terms in the response matrix  which are proportional to the overlap of the ground-state orbitals $\phi_L^0$ and $\phi_R^0$ (see Appendix~\ref{app:spec}).  Already around $b\approx 15$ states become pairwise degenerate.  
In contrast to LR-GP, the two branches of BMF excitations stay energetically well separated.

\begin{figure}[h]
           \includegraphics[width=.99\columnwidth]{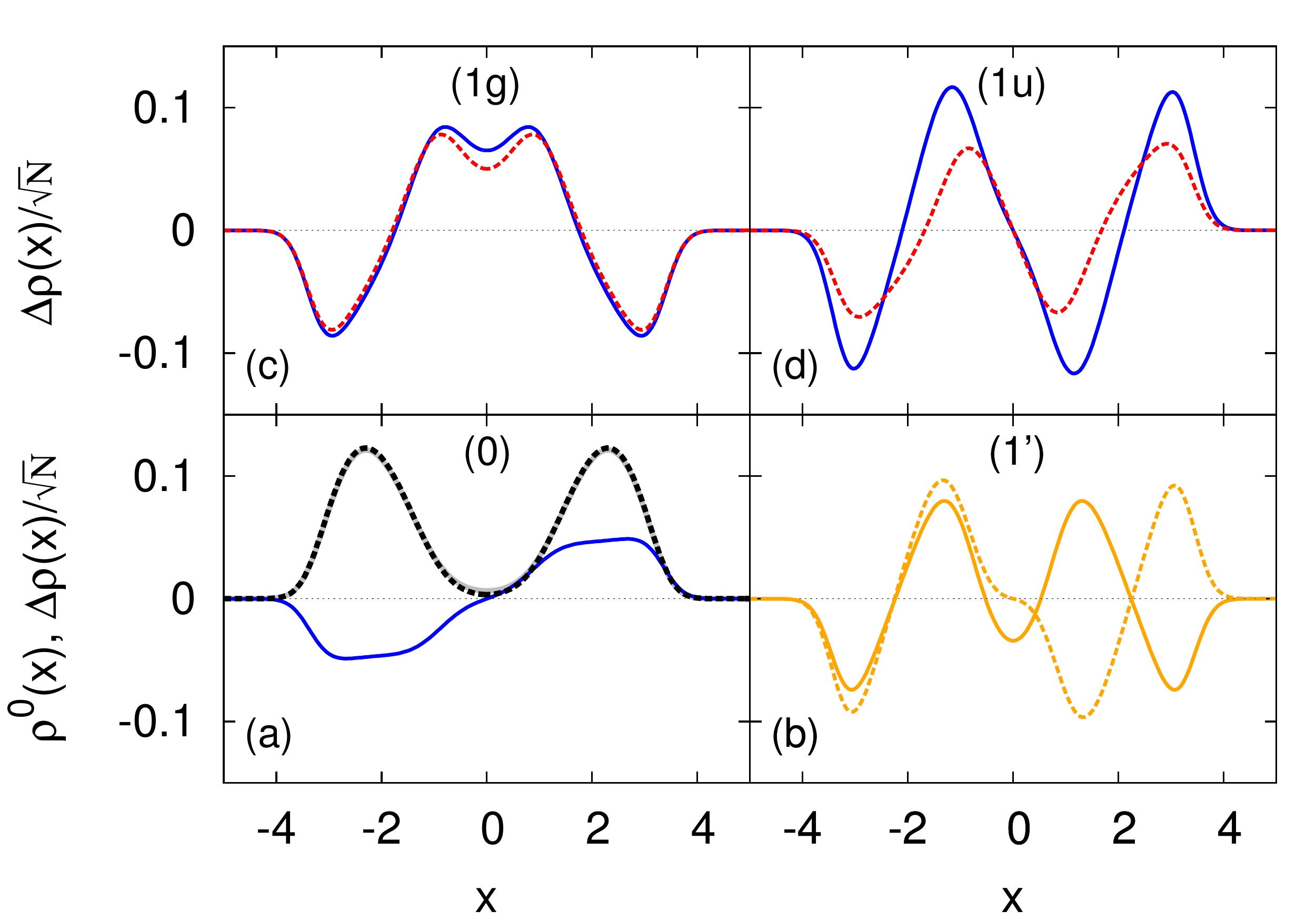}
  \caption{(Color online)  Position-space densities and density oscillations for the same double-well system as in Fig.~\ref{fig:spectrumB} for a low barrier height $b=14$. The ground state density of GP is shown in (a) by the broad gray solid line and is scaled for better comparison to the BMF density shown	 by the broad black dashed line. The GP and BMF results are seen to almost coincide. The blue solid lines in (a,c,d) show density oscillations of LR-GP. The red dashed lines in (c,d)  show density oscillations of direct excitations of LR-BMF. (a) Low-lying excitation 0 of LR-GP, which is absent for LR-BMF. (b) The gerade and ungerade swapped excitations of LR-BMF, marked as 1' and shown by the orange solid and dashed lines, respectively. (c) The gerade and (d) the ungerade excitations of LR-GP and LR-BMF, marked as  1.  See text for more details. All quantities are dimensionless.
 \label{fig:densL15}}
\end{figure}

The position-space density oscillations at low barrier $b=14$ are shown in Fig.~\ref{fig:densL15}.
\begin{figure}[h]
           \includegraphics[width=.99\columnwidth]{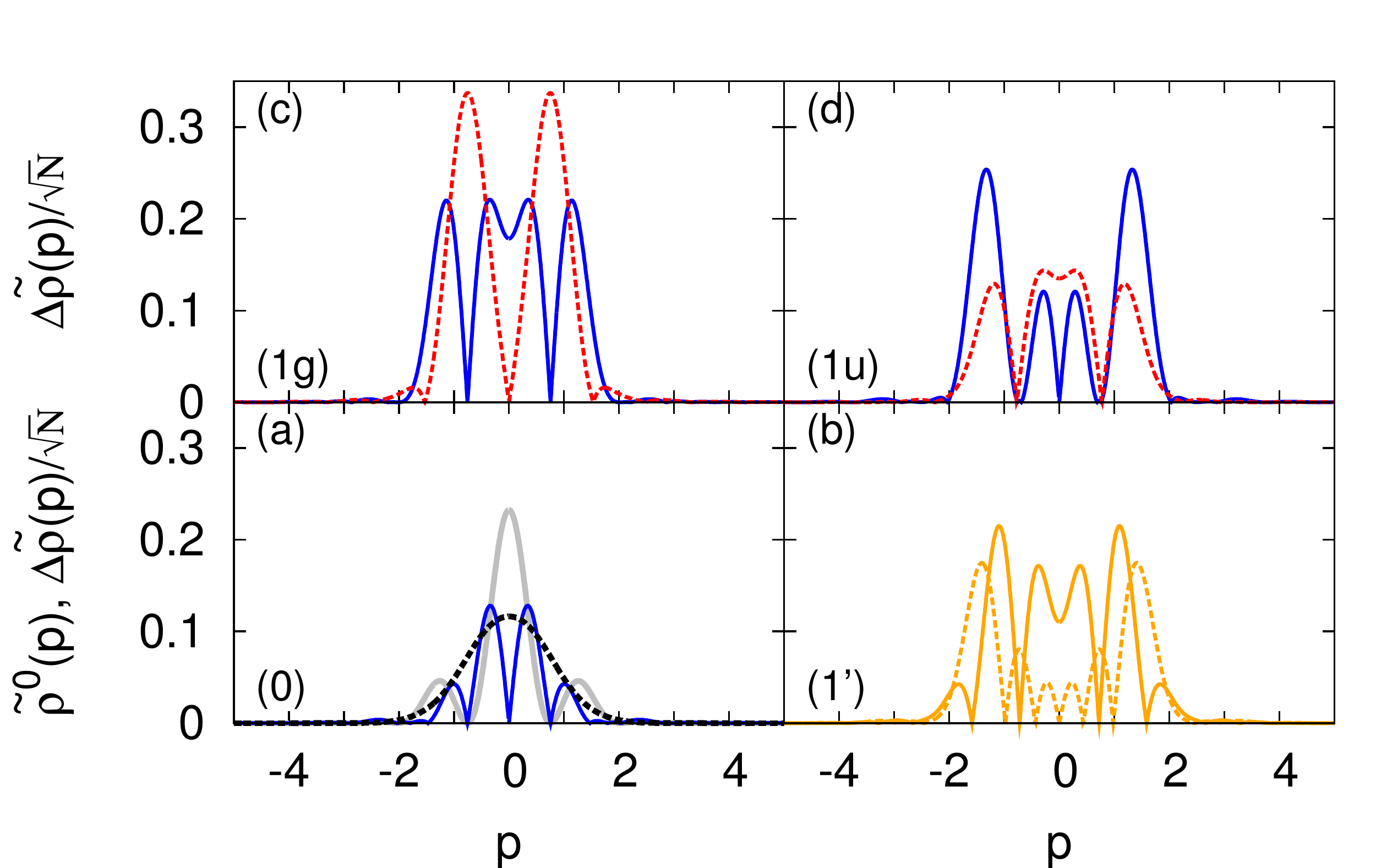}
  \caption{(Color online)  Momentum-space densities and density oscillations for the same double-well system as in Fig.~\ref{fig:spectrumB} for a low barrier height $b=14$. The ground state density of GP is shown in (a) by the broad gray solid line and is scaled for better comparison to the one of BMF, which is shown	 by the broad black dashed line. We show the same indicated excited states as in Fig.~\ref{fig:densL15}.  The blue solid lines in (a,c,d) show density oscillations of LR-GP. The red dashed lines in (c,d)  show results for direct excitations of LR-BMF. (a) Low-lying excitation 0 of LR-GP, which is absent for LR-BMF. (b) The gerade-type and ungerade-type swapped excitations of LR-BMF, marked as 1' and shown by the orange solid and dashed lines, respectively. (c) The gerade-type and (d) the ungerade-type excitations of LR-GP and LR-BMF, marked as 1.  See text for more details. All quantities are dimensionless.
 \label{fig:densL15p}}
\end{figure}
Excitation 0 as shown in (a) by the blue solid  line is qualitatively similar to the GP ground state, but it has ungerade symmetry and a flattened top. The excitations marked as 1 and 1' are shown in (c,d) and (b), respectively. Most importantly, while the gerade excitations marked as 1g are similar  for LR-GP and LR-BMF, the ungerade ones marked as 1u are quantitatively different [see (c) and (d), respectively]. Moreover, the swapped excitations of a fragmented state are approximately as large as the direct ones. Also for the density oscillations in momentum space, Fig.~\ref{fig:densL15p}, we find strong differences for condensed and fragmented systems. Hence, the lower the barrier and thus the larger the overlap of the left and right condensates, the more striking are the differences of LR-GP and LR-BMF. 

\subsection{Asymmetric double-well \label{subsec:asy}}

\begin{figure}[h]
           \includegraphics[width=.95\columnwidth]{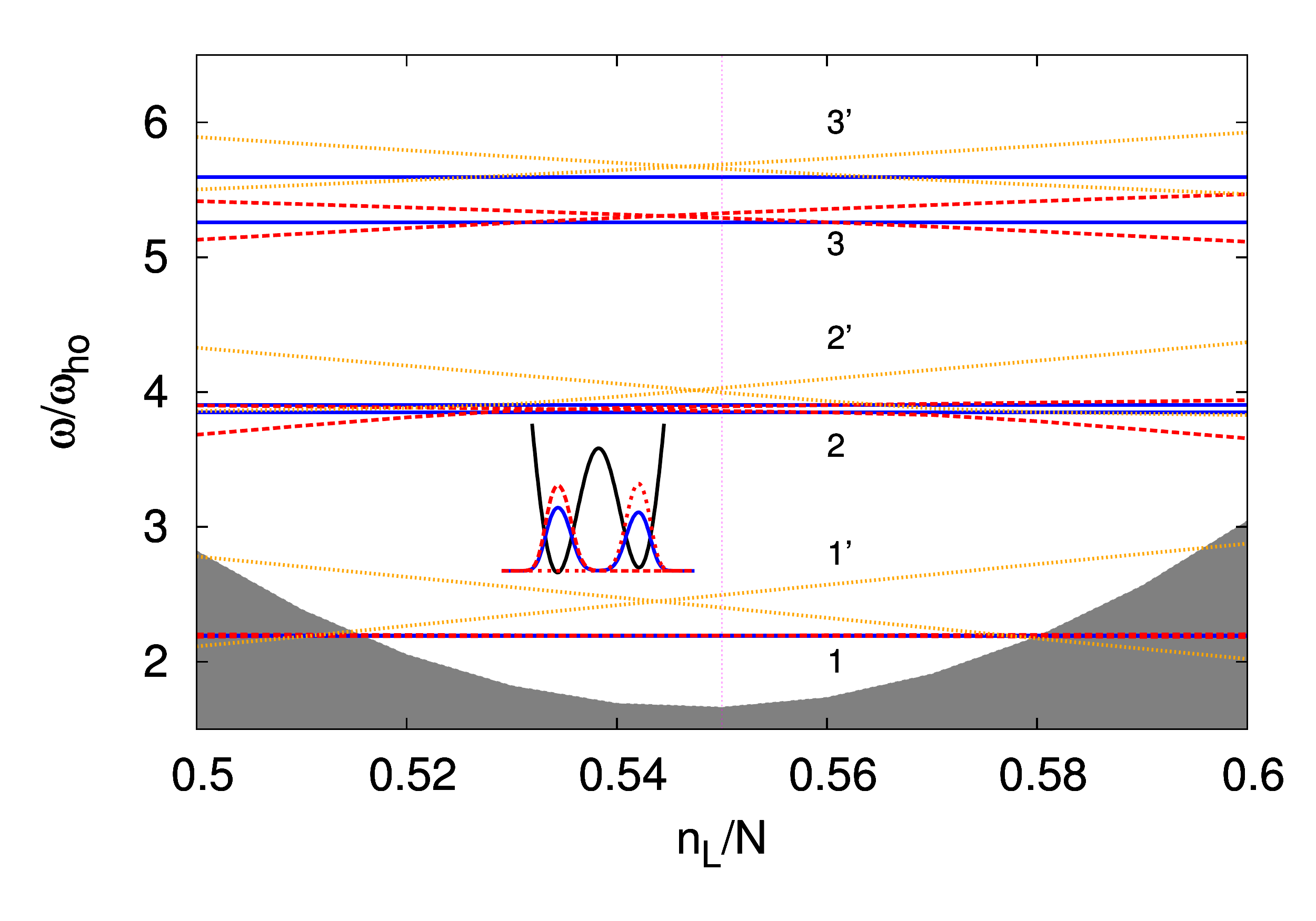}
  \caption{(Color online)  Excitation spectra of a BEC in an asymmetric double-well potential ($a=0.1$) versus occupation of the left orbital. We choose a high barrier height $b=20$ and interaction strength $\lambda_0 N=10$. The total atom number is $N=100$, and the overall harmonic confinement given by $\omega_{ho}=\sqrt{2}$. The top of the filled gray area shows the BMF ground state energies as a function of occupation of the left orbital.  The vertical dashed line marks the location of the optimal occupation $n_L/N\approx 0.55$. We compare the linear response of LR-GP, shown by the blue solid lines, and LR-BMF, shown by the red dashed lines for direct, and orange dotted lines for swapped excitations.  The excitations are grouped into pairs of lines with gerade-ungerade symmetry and marked with numbers. The swapped excitations of LR-BMF are marked with primed numbers. Inset: We plot the potential by the black  solid line. The corresponding ground state orbital  of GP at the optimal occupation is shown by the blue solid. The left (right) orbital of BMF is shown by the red dashed (short-dashed) line.  All quantities are dimensionless.
 \label{fig:spectrumN}}
\end{figure}

After having applied our response theory to BECs in a \emph{symmetric} double well potential, we now turn to a slightly \emph{asymmetric} double-well. We use an asymmetry of $a=0.1$  and choose a relatively high barrier height $b=20$ [see Eq.~\eqref{eq:pot}]. The corresponding potential is shown in the inset of Fig.~\ref{fig:spectrumN}.

\begin{figure}[h]
           \includegraphics[width=.99\columnwidth]{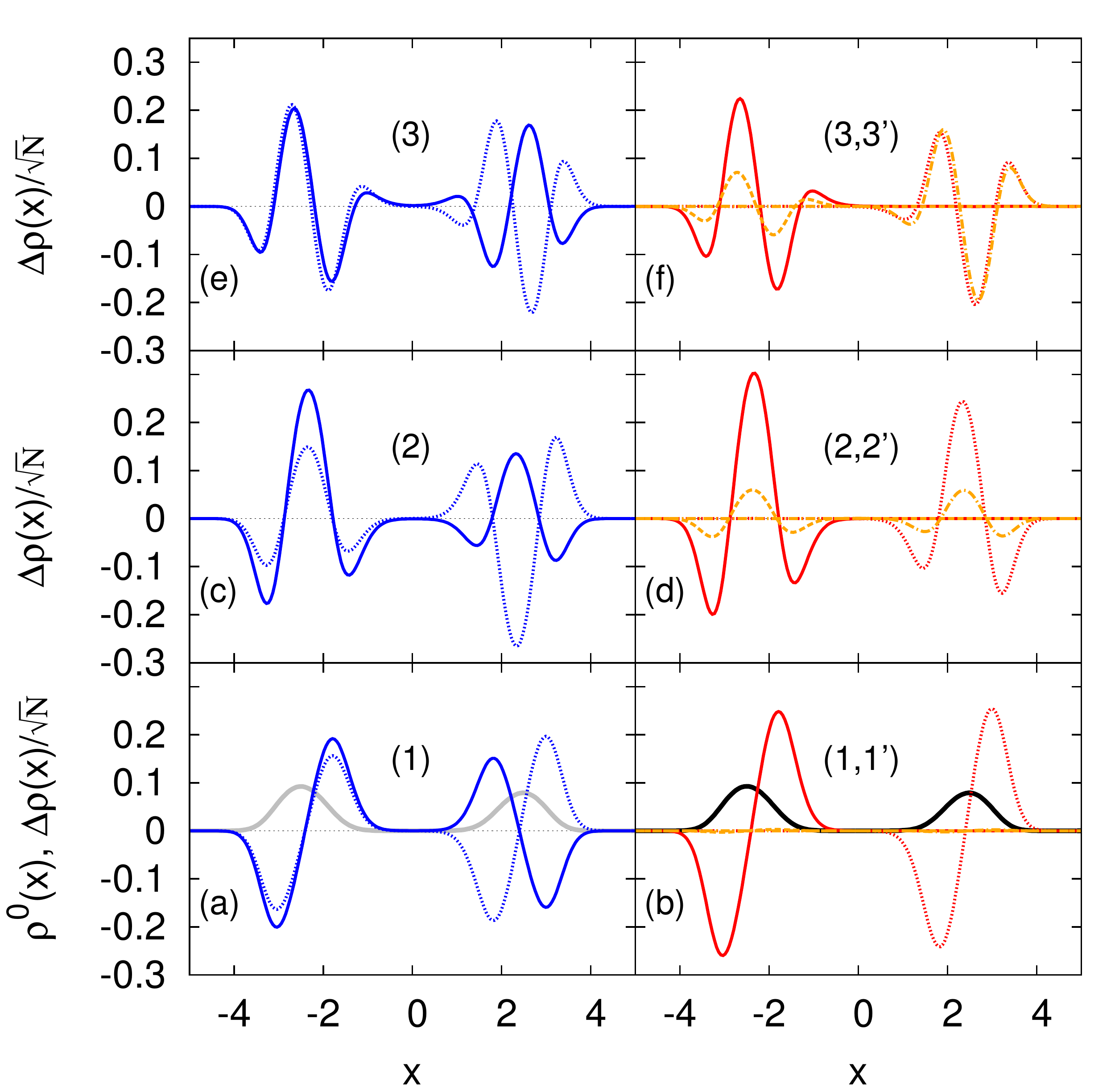}
  \caption{(Color online)  Position-space densities and density oscillations for the same double-well system as in Fig.~\ref{fig:spectrumN}  at the optimal occupation $n_L/N\approx 0.55$.
 The ground state density of GP is shown in (a) by the broad gray solid line and is scaled for better comparison to the one of BMF, which is shown	 by the broad black dashed line in (b). The density oscillations for an indicated excited state show up two lines for LR-GP (see Fig.~\ref{fig:spectrumN}). The lower in energy density oscillations are shown by the blue solid, and those higher in energy by the blue dotted lines in panels (a,c,e). A similar statement holds for the direct excitations of LR-BMF, where the respective lower excited states are shown by the red solid, and the respective upper ones by the red dotted lines in panels (b,d,f).  The lower (upper) swapped excitation are shown by the orange dashed (dashed-dotted) lines in (b,d,f). See text for more details. All quantities are dimensionless.
 \label{fig:densAsy}}
\end{figure}

Condensed and two-fold fragmented mean-field states compete for being lower in energy  also in an asymmetric double well \cite{streltsov.pra:05}. The condensed one extends over both wells and dominates in the lower well. The fragmented BEC consists of a larger fragment in the lower, and a smaller one in the upper well. We show GP (blue solid line) and BMF (red dashed lines) orbitals for typical parameters in the inset. For the chosen interaction strength $\lambda_0 N=10$ we find that the lowest in energy mean-field state is two-fold fragmented, except for very large atom numbers. In the latter case the condensed state is slightly lower in energy and the fragmented state becomes a stable excited state \cite{streltsov.pra:05}. The optimal occupation difference $n_L-n_R$ which characterizes a stable fragmented ground or excited state depends in principle on $N$. However, it takes on practically the same values already for $N\gtrsim 100$. The ground state energy versus occupation $n_L$ is shown by the filled gray line in Fig.~\ref{fig:spectrumN}, showing a minimum at $n_L/N\approx 0.55$. The ground state densities of GP and BMF, which are shown in Fig.~\ref{fig:densAsy},  perfectly coincide.

Let us first discuss how the excitation frequencies depend on the BMF occupation $(n_L,N-n_L)$ as shown in Fig.~\ref{fig:spectrumN}. Remarkably, the first excitation (marked as 1) is independent of the occupations. We can attribute this to the fact that the response amplitudes of the fragmented system are purely local in this case (not shown). They are, therefore, determined by the local ground state density, which is mostly independent of the occupations for the range of $n_L$ as shown in the figure. In contrast, the swapped excitations (marked as 1') depend linearly on $n_L$ and cross each other approximately at the optimal occupation $n_L/N=0.55$. This is due to the fact that the energy needed to excite an atom to the first excited state of the other well depends on the chemical potential difference $\mu_{LL}-\mu_{RR}$ [see also the matrix in Eq.~\eqref{eq:lrm_M2} in Appendix \ref{app:spec}]. This quantity has been identified as the energy needed to transfer a boson from one well to the other (at large enough $N$) \cite{streltsov.pra:05},
\be
\frac{d E^{M=2}}{d n_L}=\mu_{LL}-\mu_{RR}\,.
\ee
We observe that at the optimal occupation the transfer of bosons is suppressed and, therefore, $\mu_{LL}=\mu_{RR}$.  Hence, in addition to the exchange interactions also the potential difference contributes to the energetical splitting of direct and swapped excitations. The higher lying direct excitations cross each other, similar to the swapped ones. This is because as soon as the amplitudes of the direct excitations become delocalized, the energy depends on $\mu_{LL}-\mu_{RR}$. 

We next discuss the energies and the density response at the optimal BMF occupation $n_L/N=0.55$ (marked by the dotted vertical line in Fig.~\ref{fig:spectrumN}). For the lowest excitation (marked as 1) we observe that the response frequencies for both condensed and fragmented systems (direct excitations) coincide and are doubly degenerate.  The corresponding density oscillations in position space for the two solutions are shown in Fig.~\ref{fig:densAsy} (a) for LR-GP and in (b) for LR-BMF (see the solid and dashed lines, respectively). Most importantly, the density response of a simple BEC is delocalized, while  it is strictly localized for a fragmented BEC.  Also the response amplitudes are localized in the sense that either $(|u_L^k\ra,|v_L^k\ra)^T$ or $(|u_R^k\ra,|v_R^k\ra)^T$ are finite, respectively. We label the two degenerate frequencies as $k=1a,1b$. The total response at this frequency is then given by the sum of two contributions [see Eq.~\eqref{eq:density_response}], 
 proportional to
\be\label{eq:dens_sum}
\rho(x)\sim \gamma_{1a}\Delta\rho^{1a}(x)+\gamma_{1b}\Delta\rho^{1b}(x)\,.
\ee

For weak interaction strengths we can model the imbalanced ground state orbital and $u$-amplitudes for the condensed system as \footnote{The following approximations are particularly good for small $n_L-n_R$.}
\bea
&&\phi^0(x)=\left[\sqrt{n_L}\psi^0(x_L)+\sqrt{N-n_L}\psi^0(x_R)\right]/\sqrt{N}\,,\non
&&u^{1a,b}(x)=\left[\sqrt{n_L}\psi^1(x_L)\pm\sqrt{N-n_L}\psi^1(x_R)\right]/\sqrt{N}\,.
\eea
For the strictly localized orbitals and excitation amplitudes of the fragmented system we have
\bea\label{eq:asybmf}
&&\phi^0_{L}(x)=\psi^0(x_L)\,,\quad\phi^0_{R}(x)=\psi^0(x_R)\,,\non
&& u_{L}^{1a}=\psi^1(x_L)\,,\quad u_{R}^{1a}=0\,,\non
&&  u_{L}^{1b}=0\,,\quad u_{R}^{1b}=\psi^1(x_R)\,.
\eea
From this we arrive at the \emph{delocalized} density oscillations and response weights for LR-GP (assuming that the overlaps of displaced functions vanish):
\bea
&&\left[\Delta\rho^{1a,b}(x)\right]_{GP}=\left[n_L\psi^0(x_L)\psi^1(x_L)\pm(N-n_L)\psi^0(x_R)\psi^1(x_R)\right]/\sqrt{N}\,,\non
&&\gamma_{1a,b}^{GP}=\int dx f^+(x)\left[n_L\psi^0(x_L)\psi^1(x_L)\pm(N-n_L)\psi^0(x_R)\psi^1(x_R)\right]/\sqrt{N}\,.
\eea
For LR-BMF the same quantities are \emph{localized} and given as
\bea
&&\left[\Delta\rho^{1a}(x)\right]_{BMF}=\sqrt{n_L}\psi^0(x_L)\psi^1(x_L)\,,\non
&&\left[\Delta\rho^{1b}(x)\right]_{BMF}=\sqrt{N-n_L}\psi^0(x_R)\psi^1(x_R)\,,\non
&&\gamma_{1a}^{BMF}=\sqrt{n_L}\int dx f^+(x)\psi^0(x_L)\psi^1(x_L)\,,\non
&& \gamma_{1b}^{BMF}=\sqrt{N-n_L}\int dx f^+(x)\psi^0(x_R)\psi^1(x_R)\,.
\eea
The total density response of LR-GP is thus according to Eq.~\eqref{eq:dens_sum} proportional to
\be
\rho_{GP}(x)\sim 2 \left[(n_L)^2 \tilde{\gamma}_L\psi^0(x_L)\psi^1(x_L)+(N-n_L)^2\tilde{\gamma}_R\psi^0(x_R)\psi^1(x_R)\right]/N\,,
\ee
where we defined $\tilde{\gamma}_{L,R}=\int dx f^+(x)\psi^0(x_{L,R})\psi^1(x_{L,R})$ [$f^-(x)$ does not appear because the $v$-amplitudes are marginal for weak interaction strengths]. For LR-BMF we obtain
\be
\rho_{BMF}(x)\sim n_L \tilde{\gamma}_L\psi^0(x_L)\psi^1(x_L)+(N-n_L)\tilde{\gamma}_R\psi^0(x_R)\psi^1(x_R)\,.
\ee
Hence, in general the total response of the lowest excitation is different  for LR-GP and LR-BMF even for weak interaction strengths. The difference in the densities  is proportional to the imbalance $n_L-n_R$. It vanishes for symmetric occupations $n_L=N/2$. In this case, the above equations boil down to the results for a symmetric double-well potential as given in Sec.~\ref{subsec:sym}.

 For the swapped excitations the amplitudes are finite in that well where the corresponding ground state orbital vanishes [orange dashed and dashed-dotted   lines in Fig.~\ref{fig:densAsy} (b,d,f)]. Similar to the direct excitations, the position-space density response of the swapped ones is strictly localized. However, the left and right swapped excitations have different energies, and the position-space density response for the lowest two of them, marked as 1', see Fig.~\ref{fig:densAsy} (b) (orange dashed and dashed-dotted lines), is very small. 

\begin{figure}[h]
           \includegraphics[width=.99\columnwidth]{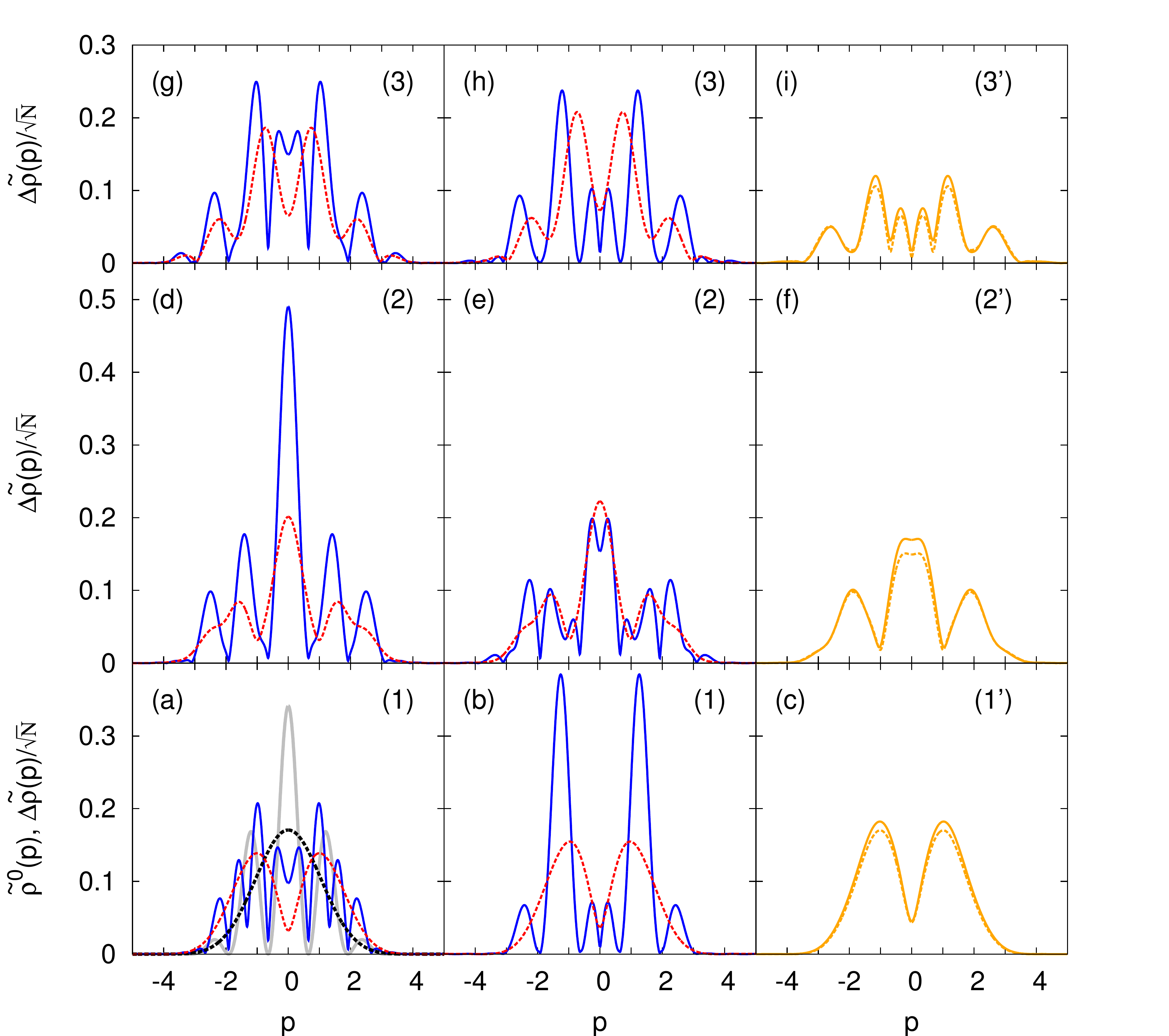}
  \caption{(Color online) Momentum-space densities and density oscillations  for the same double-well system as in Fig.~\ref{fig:spectrumN} at the optimal occupation $n_L/N\approx 0.55$. The ground state density of GP is shown in (a) by the broad gray solid line and is scaled for better comparison to the one of BMF, which is shown	 by the broad black dashed line. We show the same indicated excitations as in Fig.~\ref{fig:densAsy}.  The density oscillations for a given number show up two lines for LR-GP (see Fig.~\ref{fig:spectrumN}). The lower in energy density oscillations are shown by the blue solid lines in (a,d,g), and those higher in energy by the blue dotted lines (b,e,h). A similar statement holds for the direct excitations of LR-BMF, where the respective lower excited states are shown by the red dashed lines in (a,d,g), and the respective upper ones in (b,e,h).  The lower (upper) swapped excitations of LR-BMF are shown by the orange solid (dashed) lines in (c,f,i).    See text for more details. All quantities are dimensionless.
 \label{fig:denspAsy}}
\end{figure}

We next discuss the density response in momentum space,    see Fig.~\ref{fig:denspAsy} (a-c).
For LR-GP (blue solid lines) it is similar as in the symmetric double-well [compare with the respective results plotted in Fig.~\ref{fig:densp20} (a-c)]. For LR-BMF, however, the momentum-space density oscillations are very different as compared to the symmetric case -- they have very little structure and only one node. This can be explained by the local nature of the excitations. For weak interaction strengths the momentum-space density oscillations of the direct as well as the swapped excitations are, following from Eq.~\eqref{eq:asybmf}, just given by
\bea\label{eq:p_phase_a}
&&\left|[\Delta\tilde\rho^{k}(p)]_{BMF}\right|=\sqrt{n_L}\tilde{\psi}^0(p)\left|\tilde{\psi}^1(p)\right|\,,\quad\mathrm{for}\quad k=1a,1'b\,,\non
&&\left|[\Delta\tilde\rho^{k}(p)]_{BMF}\right|=\sqrt{N-n_L}\tilde{\psi}^0(p)\left|\tilde{\psi}^1(p)\right|\,,\quad\mathrm{for}\quad k=1b,1'a\,.
\eea
Hence, the  phase factors which have been found in case of a  symmetric double-well [Eqs.~\eqref{eq:p_phase}], and which are due to an interference of the \emph{left} and the \emph{right} response, are absent here. 

For the higher excitations we find an energetical splitting of the lines, see Fig.~\ref{fig:spectrumN}. For LR-GP the density oscillations with an imbalance on the left (blue solid lines in  Fig.~\ref{fig:densAsy}) or right (blue dotted lines) occur at different frequencies. Similarly, the left (red solid and orange dashed lines) and right (red dotted and orange dashed-dotted lines) localized density oscillations of LR-BMF become energetically separated. We note that the shapes of the density oscillations are strongly affected by the contribution of the off-diagonal Lagrange multipliers $\mu_{LR}$ and $\mu_{RL}$ in the linear response equation [see Eq.~\eqref{eq:uL}]. 
Hence, in contrast to a condensed state, \emph{the position-space density response of a fragmented state is purely local with generally different frequencies and density oscillations for the left and right response.}

\section{Summary and conclusions \label{sec:con}}

The Bogoliubov-de Gennes equations are the standard linear response equations for bosons. They are
applicable for condensates where all atoms occupy only a single orbital. We presented the
linear response theory for fragmented condensates, where the atoms are allowed to be
distributed over several orbitals. We derived the linear response equations for
a small periodic perturbation of a stationary state. Since our linear response theory is based
on the BMF and TDMF methods, we call the derived equations LR-BMF. Those allow us
to determine excitation energies and response amplitudes of fragmented condensates with
an arbitrary degree of fragmentation.

We analyzed the properties of LR-BMF. Most notably, in the derived equations the
response of each fragment is orthogonal to all the ground-state orbitals. This has vast
consequences on the shape and the energies of the excitations. The linear response matrix
defines a biorthogonal basis, which consists of the vector of response amplitudes related to
all the fragments. The response of the fragments is coupled through the Lagrange multipliers, and whenever they overlap in space.
The Lagrange multipliers of the BMF ground-state orbitals enter the linear response matrix
and account for the relative energies of the stationary orbitals. We give expressions for the
density oscillations in real and momentum space which arise due to a resonant perturbation.
They are given as sums of the contributions due to the response of each fragment, weighted
by the square root of the orbitals' occupations.

As applications, we investigated Bose-Einstein condensates in symmetric as well as
asymmetric double-well potentials. We compare results of the LR-GP (i.e., Bogoliubov-de
Gennes) and LR-BMF theories. Our numerical
results demonstrate that the responses of a fragmented and a condensed system show striking
differences. In particular, fragmented BECs possess a class of swapped excitations which
do not exist in condensed systems. They are characterized by a response which is dominantly
in the respective neighboring well. These excitations signify the transition from below the
barrier to above the barrier excitations. The density response in momentum space has been
found to be qualitatively different, even in situations where the response energies of the two
theories numerically coincide. At low barrier heights, an excitation to the ungerade state of GP exists within LR-GP, but it is absent within LR-BMF. Thus it can be used as a signature of coherence. For fragmented BECs in asymmetric double-well potentials we found a localized
density response (i.e., finite in either one or the other well), as well as an energy splitting between left and right response. This is in
stark contrast to the response of a condensate which is not fragmented.

We conclude that, for a proper analysis of the response of even one-body observables like
density, it is crucial to take into account the many-body structure of the underlying state. In
view of the rich physics which has been found using the standard Bogoliubov-de Gennes
equations, our generalization of this very successful theory to fragmented BECs offers even
more rich prospects for understanding excitations of cold atom systems in general. The vast
differences between the response of condensed and fragmented systems will provide a way to
distinguish condensed and fragmented states by linear response experiments.

\section*{Acknowledgements}

We thank A. U. J. Lode and K. Sakmann for valuable discussions and for advises in numerical issues. JG appreciates support from the Alexander von Humboldt Foundation. Financial support by the DFG and the STREP project `QIBEC' are gratefully acknowledged.

\begin{appendix}

\section{Linear response of TDMF without constraint \label{app:lin}}

We discuss here shortly the derivation of linear response of the full form of the TDMF equations [see Eq.~\eqref{eq:TDMFP}]. The projector on the left hand side of Eq.~\eqref{eq:TDMFP} translates to a projector on the term proportional to $\omega$ in the linear response equations [see Eq.~\eqref{eq:lr}]. As a consequence, the response amplitudes $\left(|\mathbf u\ra,|\mathbf v\ra\right)^T$ are not necessarily orthogonal to the ground-state orbitals $\boldsymbol{\phi^0}$ [i.e., Eq.~\eqref{eq:condP} does not hold]. Thus, the question arises if in addition to the orthogonal eigenvectors as defined in Eq.~\eqref{eq:evL}, also the ground-state orbitals have to be included into the ansatz of the response amplitudes in Eq.~\eqref{eq:ansatz1}. Since we expanded around stationary BMF (ground state) orbitals $\boldsymbol{\phi^0}$, we assume that they are recovered if the perturbation is zero (i.e., $f^+=f^-=0$). 
Thus, the response amplitudes $\left(|\mathbf u\ra,|\mathbf v\ra\right)^T$ can contain solely those ground state orbital contributions, which lead to trivial time-dependent phases on the orbitals. We note that those are determined from BMF only modulo a phase. Thus, also when we start with the full form of the TDMF equations [see Eq.~\eqref{eq:TDMFP}], the frequencies $\omega_k$ (excitation spectra), the response amplitudes $|\mus^k\ra$ and $|\mvs^k\ra$, as well as the perturbed orbitals $\boldsymbol{\phi}_k$ are the same as for the linear response of the TDMF working equations [see Eq.~\eqref{eq:TDMF}].

\section{Special cases of linear response: $M=1$ and $M=2$ \label{app:spec}}

\subsubsection*{Linear response of a condensed state ($M=1$)}

For $M=1$ we recover the results for the excitation spectra as obtained from the number-conserving Bogoliubov theory of Ref.~\cite{castin:98}, with the linear response matrix from Eq.~\eqref{eq:lrm}
\be\label{eq_app:lrm_gp}
\boldsymbol{\mathcal{L}_{M=1}}=\left(\begin{array}{cc} \hat H_{GP}^0 +\hat P\lambda|\phi^0|^2\hat P-\mu & \hat P\lambda(\phi^0)^2\hat P^*\\
-\hat P^*\lambda(\phi^{0,*})^2\hat P & -\hat H_{GP}^0 -\hat P^*\lambda|\phi^0|^2\hat P^*+\mu
 \end{array}\right)\,.
\ee
We stress that in our derivation the orthogonality of the response amplitudes $(|u\ra,|v\ra)^T$ to the ground-state orbitals $\phi^0$ is obtained naturally from the derivation, in contrast to Ref.~\cite{castin:98} where it is an assumption.

Interestingly, the standard and the number-conserving BdG equations can be considered as the linear response of different forms of the GP equation, which deviate by a global phase on $\phi(\mr,t)$ \cite{alon.pla:07}. Such a phase has no physical meaning, and, therefore, the three forms of the GP equation are equivalent and predict the same physics. The  standard form, with response matrix given by Eq.~\eqref{eq:lrm_gp}, is obtained from the usual form of the GP equation:
\be\label{eq:GP}
i\dot{\phi}=\hat H_{GP}\phi\;.
\ee
If we consider the TDMF equations [see Eq.~\eqref{eq:TDMF}] for the case of $M=1$, we obtain a GP equation which is similar to that obtained from a number-conserving approach (see Ref.~\cite{castin:98}):
\be\label{eq:GP_P}
i\dot{\phi}=\hat P\hat H_{GP}\phi\,.
\ee
The linear response of this equation leads exactly to the linear response matrix of Eq.~\eqref{eq_app:lrm_gp}. Both the standard and the number-conserving BdG equations lead to the same response frequencies $\omega^k$. The response amplitudes in the number-conserving formalism are obtained from those of the standard form by projecting into the subspace orthogonal to the ground state orbital $\phi^0(\mr)$ with the projector $\hat P$ 
\cite{castin:98}. The difference in the perturbed orbitals $\phi(\mr,t)$ is then just a trivial phase. However, the zero eigenvectors of both forms differ from each other. In particular, in the standard, number-non-conserving linear response there is one missing zero eigenvector, which is supposed to lead to a divergence of quantum fluctuations in time \cite{lewenstein:96,castin:98}. 

The full form of the TDMF equations [see Eq.~\eqref{eq:TDMFP}], including the projector on the left hand side, reads
\be
\hat P i \dot{\phi}=\hat P\hat H_{GP}\phi\,,
\ee
having the same linear response as Eq.~\eqref{eq:GP_P}, apart from contributions of the ground state orbital to the solution of the perturbed orbital [Eq.~\eqref{eq:response_gp}], which corresponds to a physically irrelevant global phase.

\subsubsection*{Linear response of a two-fold fragmented state ($M=2$)}

We explicitly write down the linear response matrix which is used in the application part, Sec.~\ref{sec:app}, as a special case ($M=2$) of the general formula, Eq.~\eqref{eq:lrm}. Since the orbitals of a two-fold fragmented BEC in a double-well potential are localized, we use as orbitals' indices left `L' and right `R'. We obtain as the linear response matrix $\boldsymbol{\mathcal{L}_{M=2}}=\boldsymbol{\mathcal{P}}\boldsymbol{\mathcal{L}^{'}_{M=2}}\boldsymbol{\mathcal{P}}$:
\begin{widetext}

\bea\label{eq:lrm_M2}
&&\boldsymbol{\mathcal{L}^{'}_{M=2}}=\\
&&\left(\begin{array}{l|l|l|l} \hat Z^0_L+\tilde{n}_L|\phi_L^0|^2-\mu_{LL}^0&2 \bar{n}\phi_R^{0,*}\phi_L^0-\sqrt{\frac{\tilde{n}_L}{\tilde{n}_R}}\mu_{LR}^0&\tilde n_L\left(\phi_L^0\right)^2&2\bar{n}\phi_R^0\phi_L^0\\
\hline 2\bar{n}\phi_L^{0,*}\phi_R^0-\sqrt{\frac{\tilde{n}_R}{\tilde{n}_L}}\mu_{RL}^0&\hat Z^0_R+\tilde n_R|\phi_R^0|^2-\mu_{RR}^0&2\bar{n}\phi_L^0\phi_R^0&\tilde n_R(\phi_R^0)^2\\
\hline -\tilde n_L\left(\phi_L^{0,*}\right)^2&-2\bar{n}\phi_R^{0,*}\phi_L^{0,*}&-\hat Z^0_L-\tilde n_L|\phi_L^0|^2+\mu_{LL}^{0,*}&-2\bar{n}\phi_L^{0,*}\phi_R^0+\sqrt{\frac{\tilde{n}_L}{\tilde{n}_R}}\mu_{LR}^{0,*}\\
\hline -2\bar{n}\phi_L^{0,*}\phi_R^{0,*}&-\tilde n_R\left(\phi_R^{0,*}\right)^2&-2\bar{n}\phi_L^0\phi_R^{0,*}+\sqrt{\frac{\tilde{n}_R}{\tilde{n}_L}}\mu_{RL}^{0,*}&-\hat Z^0_R-\tilde n_R|\phi_R^0|^2+\mu_{RR}^{0,*} \end{array}\right)\,,\nonu
\eea
 
\end{widetext}
where we use the notations $\tilde n_i=\lambda_0(n_i-1)\approx\lambda_0 n_i$, and $\bar{n}=\lambda_0\sqrt{n_L n_R}$. The latter approximation is only chosen because it makes the linear response matrix appear simpler. The numerics in this paper are performed exactly, i.e., without this approximation. We divided the matrix into blocks. For very weak interaction strengths, the $v$-amplitudes are zero and the $u$-amplitudes are then solely determined by the upper left $2\times2$-block. The diagonals of this submatrix account for the excitation energy contributions due to the external and interaction potential of the corresponding fragment. The off-diagonals account for the coupling energy to the other fragment. For stronger interaction strengths, the off-diagonal $2\times2$-blocks become important and induce finite $v$-amplitudes. As we have seen in Sec.~\ref{sec:app}, those lead for example to the damping of density oscillations. For spatially disjunct orbitals, i.e.,  $\int dx |\phi_L^0(x)||\phi_R^0(x)|=0$, the linear response matrix boils down to two independent matrices, each acting on a separate subsystem.

We note that unlike the case $M=1$, where the eigenvectors of $\boldsymbol{\mathcal{P}}\boldsymbol{\mathcal{L}}\boldsymbol{\mathcal{P}}$ can be obtained   by application of the projection operator on the eigenvectors of $\boldsymbol{\mathcal{L}}$ \cite{castin:98}, this does not hold anymore for fragmented states\footnote{We note that it also does not hold for the multi-component GP equation, see Appendix~\ref{app:com}.}, i.e., for $M>1$. We demonstrate this property for $M=2$. The eigenvectors of $\boldsymbol{\mathcal{L}^{'}_{M=2}}$, given by 
\be
\boldsymbol{\mathcal{L}^{'}_{M=2}}\left(\begin{array}{c}|U_L^k\ra\\|U_R^k\ra\\|V_L^k\ra\\|V_R^k\ra\end{array}\right)=\tilde{\omega}^k\left(\begin{array}{c}|U_L^k\ra\\|U_R^k\ra\\|V_L^k\ra\\|V_R^k\ra\end{array}\right)\,,
\ee
are in general not orthogonal to the ground-state orbitals. For the statement to be valid, the following expression must vanish:
\be\label{eq_app:cond}
\boldsymbol{\mathcal{P}}\boldsymbol{\mathcal{L}^{'}_{M=2}}(1-\boldsymbol{\mathcal{P}})\left(\begin{array}{c}|U_L^k\ra\\|U_R^k\ra\\|V_L^k\ra\\|V_R^k\ra\end{array}\right)=\boldsymbol{\mathcal{P}}\boldsymbol{\mathcal{L}^{'}_{M=2}}\left(\begin{array}{c}|\phi_L^0\ra\la\phi_L^0|U_L^k\ra+|\phi_R^0\ra\la\phi_R^0|U_L^k\ra\\|\phi_L^0\ra\la\phi_L^0|U_R^k\ra+|\phi_R^0\ra\la\phi_R^0|U_R^k\ra\\|\phi_L^{0,*}\ra\la\phi_L^{0,*}|V_L^k\ra+|\phi_R^{0,*}\ra\la\phi_R^{0,*}|V_L^k\ra\\|\phi_L^{0,*}\ra\la\phi_L^{0,*}|V_R^k\ra+|\phi_R^{0,*}\ra\la\phi_R^{0,*}|V_R^k\ra \end{array}\right)\,.
\ee
This is not the case since the only eigenvector of $\boldsymbol{\mathcal{P}}\boldsymbol{\mathcal{L}^{'}_{M=2}}$ with eigenvalue zero which lies in the space spanned by the ground state vectors is $(|\phi_L^0\ra,|\phi_R^0\ra,|\phi_L^{0,*}\ra,|\phi_R^{0,*}\ra)^T$.\footnote{The other eigenvectors with eigenvalue zero can be constructed as vectors which are transformed by $\boldsymbol{\mathcal{L}^{'}_{M=2}}$ into a linear combination of ground-state orbitals, similar as for BdG in Ref.~\cite{castin:98}.}
As stated above, the only exception is $M=1$, where generally $|u\ra=|\hat PU\ra$ and $|v\ra=|\hat PV\ra$ hold. Then, by using the orthonormalization relations Eq.~\eqref{eq:orth} for $M=1$, we find that the vector appearing on the right hand side of Eq.~\eqref{eq_app:cond} turns out to be proportional to the zero-eigenvector $(|\phi^0\ra,|\phi^{0,*}\ra^T)$ of $\boldsymbol{\mathcal{L}_{BdG}}$ [see Eq.~\eqref{eq:lrm_gp}] \cite{castin:98}.

\section{Comparison to the linear response of two-component GP equations \label{app:com}}

In this Appendix we compare the linear response of a two-fold fragmented condensate derived here, see Appendix~\ref{app:spec}, to that of a two-component BEC, see for example Ref.~\cite{Pu:98}. The latter system is described by two coupled Gross-Pitaevskii (2GP) equations \cite{Pu:98,Esry.pra:98,Esry.prl:97,sorensen:02}
\bea\label{eq_app:2GP}
&&i\dot{\phi}_L^0(x)=\left\{\hat h_L(x)+\lambda_{LL}(n_L-1)|\phi_L^0(x)|^2+\lambda_{LR} n_R|\phi_R^0(x)|^2\right\}\phi_L^0(x)\,,\non
&&i\dot{\phi}_R^0(x)=\left\{\hat h_R(x)+\lambda_{RR}(n_R-1)|\phi_R^0(x)|^2+\lambda_{LR} n_L|\phi_L^0(x)|^2\right\}\phi_R^0(x)\,.
\eea
We borrow the ``above" notation with `L' and `R' representing the two species.
For each component, the single-particle Hamiltonian is given by $h_{L,R}(x)=-\frac{1}{2 m_{L,R}}\frac{\partial^2}{\partial x^2}+V_{L,R}(x)$, taking into account the different mass and potential trap of each component. The equations are different from the TDMF equations [see Eq.~\eqref{eq:TDMF}] in two respects. Firstly, the two-component GP equations~\eqref{eq_app:2GP} do not contain projectors $\hat P$, since the atoms in different components are distinguishable. Secondly, the factor of 2 which appears in the TDMF equations due to the exchange interactions between identical particles is absent here. Instead, for two component BECs we have three interaction parameters, where $\lambda_{LL}$ and $\lambda_{RR}$ account for the interactions between atoms of the same species, and $\lambda_{LR}$ between atoms of different species. A dynamical comparison of single-component fragmented and two species BECs, examining the case of interaction-assisted self interference in free space, can be found in Ref.~\cite{cederbaum.prl:07}. 

Those differences between the identical and distinguishable particles  also translate to the linear response of the two-orbital TDMF and the 2GP equations. In particular, we have found that for fragmented single-species BECs the response is orthogonal to \emph{all} of the ground-state orbitals. This is not, of course, the case with two coupled GP equations, where, even when one employs a number-conserving framework for linear response as in Ref.~\cite{castin:98}, the excitations of a given species are orthogonal solely to the ground state of the same species. This becomes important, e.g., for two-component BECs in a double-well potential, where at sufficiently high barrier each component is localized in one well. Then the excitation of the left species in the right well can have the same (say Gaussian) shaped ground state as the other species. For identical bosons we found a different behavior, where all excitations had at least one node in order to ensure orthogonality, see Sec.~\ref{sec:app}. Another property of the two-component case is that one can investigate observables depending only on one orbital (species) \cite{Esry.pra:98}, in contrast to identical atoms where this is not possible. 

\section{Linear response of the Bose-Hubbard model \label{app:lr_bh}}

The assumption underlying BMF is that the ground state of the double-well is a perfect mean-field fragmented state, i.e. a Fock state [see Eq.~\eqref{eq:MFansatzBMF}]. We investigate here the effects of a small tunnel coupling due to overlapping BMF orbitals by employing the two-site Bose-Hubbard model \cite{jaksch:98,grond.pra:09b} in order to describe the hopping between the two BMF orbitals $\phi_L^0(x)$ and $\phi_R^0(x)$. 
The Hamiltonian is given by
\begin{equation}\label{eq:hamtwomode}
  \hat H^{BH}=-\frac{\Omega(t)}2\left(\hat a_L^\dagger \hat a_R^{\phantom\dagger}+
  \hat a_R^\dagger \hat a_L^{\phantom\dagger}\right)+E\left(
  \hat a_L^\dagger\hat a_L -
  \hat a_R^\dagger\hat a_R
  \right)
  +
  \kappa\left(
  \hat a_L^\dagger\hat a_L^\dagger\hat a_L^{\phantom\dagger}\hat a_L^{\phantom\dagger}+
  \hat a_R^\dagger\hat a_R^\dagger\hat a_R^{\phantom\dagger}\hat a_R^{\phantom\dagger}
  \right)\,,
\end{equation}
where $a_{L}^\dagger$ and $a_{R}^\dagger$ create an atom in the left and right localized orbitals, respectively. The tunnel coupling is given by $\Omega(t)=2\int dx\phi_L^0(x)\hat h(x,t)\phi_R^0(x)$, the asymmetry by $ E=\int dx\phi_L^0(x)\hat h(x)\phi_L^0(x)-\int dx\phi_R^0(x)\hat h(x)\phi_R^0(x)$. ${\kappa}=\frac{\lambda_0}{2}\int dx|\phi_{L}^0(x)|^4\approx \frac{\lambda_0}{2}\int dx|\phi_{R}^0(x)|^4$
is the prefactor of the interaction term. $\hat H^{BH}$ acts on an $(N+1)$-dimensional state vector $\mathbf{C}$, which entries mark the probabilities for having $n$ atoms in the left and $N-n$ in the right orbital.

In order to get the linear response of the BH model, which we call LR-BH, we employ a time-dependent perturbation of the external potential as $\delta\hat h=f(x)\sin{(\omega t)}$. Since the resonance frequencies of the orbitals as obtained from LR-BMF are in general different than the energies for hopping, i.e., the resonance frequencies of LR-BH, we assume time-independent orbitals $\phi_L^0(x)$ and $\phi_R^0(x)$. We note that a general study of the interplay between orbitals' and hopping excitations requires a full many-body analysis.

The perturbation affects the tunnel coupling: $\Omega(t)\rightarrow\Omega(t)+\delta\Omega(t)$, with
\be
\delta\Omega(t)=2\sin{(\omega t)}\int dx\phi_L^0(x)f(x)\phi_R^0(x)\,,
\ee
as well as the energy difference between left and right states
\be
\delta E(t)=\sin{(\omega t)}\int dx\left(|\phi_L^0(x)|^2-|\phi_R^0(x)|^2\right)f(x)\,.
\ee
Linearizing $\mathbf{C}\approx \mathbf{C}^0+\delta\mathbf{C}$, we obtain
\be
i\dot{\delta\mathbf{C}}=\hat H^{BH}\delta\mathbf{C}+\left[\delta\Omega(t)\left(\hat a_L^\dagger \hat a_R^{\phantom\dagger}+
  \hat a_R^\dagger \hat a_L^{\phantom\dagger}\right)+\delta E(t)\left(\hat a_L^\dagger \hat a_L^{\phantom\dagger}-
  \hat a_R^\dagger \hat a_R^{\phantom\dagger}\right)\right]\mathbf{C}^0/2\,,
\ee
which can be straightforwardly solved by 
\be
\delta\mathbf{C}=-\sin{(\omega t)}\sum_k\gamma_k \mathbf{C}^k/(\omega-\omega_k)\,.
\ee
Hereby, $\mathbf{C}^k$ and $\omega_k$ correspond simply to the eigenstates and eigenfrequencies of $\hat H^{BH}$, respectively. More importantly, the response weights  $\gamma_k=\gamma_k^{\Omega}+\gamma_k^{E}$ are given by 
\be
\gamma_k^{\Omega}=\la \mathbf{C}^k|\hat a_L^\dagger \hat a_R^{\phantom\dagger}+
  \hat a_R^\dagger \hat a_L^{\phantom\dagger}|\mathbf{C}^0\ra/2 \int dx \phi_L^0(x) f(x) \phi_R^0(x)\,,
\ee
and
\be
\gamma_k^{E}=\la \mathbf{C}^k|\hat a_L^\dagger \hat a_L^{\phantom\dagger}-
  \hat a_R^\dagger \hat a_R^{\phantom\dagger}|\mathbf{C}^0\ra/2 \int dx \left(|\phi_L^0(x)|^2 - |\phi_R^0(x)|^2\right) f(x) \,.
\ee
For the response weights related to tunneling, which are proportional to the overlap of the orbitals, we find for all setups discussed in this paper $\gamma_k^{\Omega}\approx 0$. For the response weights related to the potential asymmetry, we observe that in a symmetric double-well $\gamma_k^{E}=0$ for symmetric perturbations $f(x)$. For asymmetric perturbations, $\gamma_k^{E}$ is non-negligible only for a few low lying excitations of LR-BH, since it scales as $\sim\la \mathbf{C}^k|\hat a_L^\dagger \hat a_L^{\phantom\dagger}-
  \hat a_R^\dagger \hat a_R^{\phantom\dagger}|\mathbf{C}^0\ra$. For example, we checked that for $N=100$ only the lowest excitation of LR-BH gives rise to a nonzero response. Moreover, one can show that the total position-space density response [i.e., $\rho(x,t)$ as in Eq.~\eqref{eq:density_response}] at the LR-BH resonance frequencies scales with $\sim\left(\la \mathbf{C}^k|\hat a_L^\dagger \hat a_L^{\phantom\dagger}-
  \hat a_R^\dagger \hat a_R^{\phantom\dagger}|\mathbf{C}^0\ra\right)^2$ and is thus marginal.  

\end{appendix}


%

\end{document}